\documentclass[english,preprintnumbers,amsmath,amssymb,pr,nofootinbib,notitlepage]{revtex4-1}
\usepackage[latin9]{inputenc}
\usepackage{color}
\usepackage{amsmath}
\usepackage{amssymb}
\usepackage{graphicx}

\makeatletter
\numberwithin{equation}{section}

\usepackage{epsfig}

\usepackage{dcolumn}

\makeatother

\usepackage{babel}
\begin{document}
\title{Quantum fluxes at the inner horizon of a near-extremal spherical charged
black hole}
\author{Noa Zilberman}
\email{noazilber@campus.technion.ac.il}

\author{Amos Ori}
\email{amos@physics.technion.ac.il}

\affiliation{Department of Physics, Technion, Haifa 32000, Israel}
\date{\today}
\begin{abstract}
We analyze and compute the semiclassical stress-energy flux components,
the outflux $\left\langle T_{uu}\right\rangle _{\text{ren}}$ and
the influx $\left\langle T_{vv}\right\rangle _{\text{ren}}$ ($u$
and $v$ being the standard null Eddington coordinates), at the inner
horizon (IH) of a Reissner-Nordström black hole (BH) of mass $M$
and charge $Q$, in the near-extremal domain in which $Q/M$ approaches
$1$. We consider a minimally-coupled massless quantum scalar field,
in both Hartle-Hawking and Unruh states, the latter corresponding
to an evaporating BH. The near-extremal domain lends itself to an
analytical treatment which sheds light on the behavior of various
quantities on approaching extremality. We explore the behavior of
the three near-IH flux quantities $\left\langle T_{uu}^{-}\right\rangle _{\text{ren}}^{U}$,
$\left\langle T_{vv}^{-}\right\rangle _{\text{ren}}^{U}$, and $\left\langle T_{uu}^{-}\right\rangle _{\text{ren}}^{H}=\left\langle T_{vv}^{-}\right\rangle _{\text{ren}}^{H}$,
as a function of the small parameter $\Delta\equiv\sqrt{1-\left(Q/M\right)^{2}}$
(where the superscript ``$U$'' or ``$H$'' respectively refers
to the Unruh or Hartle-Hawking state and  ``$-$'' refers to the
IH value). We find that in the near-extremal domain $\left\langle T_{uu}^{-}\right\rangle _{\text{ren}}^{U}\cong\left\langle T_{uu}^{-}\right\rangle _{\text{ren}}^{H}=\left\langle T_{vv}^{-}\right\rangle _{\text{ren}}^{H}$
behaves as $\propto\Delta^{5}$. In contrast, $\left\langle T_{vv}^{-}\right\rangle _{\text{ren}}^{U}$
behaves as $\propto\Delta^{4}$, and we calculate the prefactor analytically.
It therefore follows that the semiclassical fluxes at the IH neighborhood
of an evaporating near-extremal spherical charged BH are dominated
by the influx $\left\langle T_{vv}\right\rangle _{\text{ren}}^{U}$. 

In passing, we also find an analytical expression for the transmission
coefficient outside a Reissner-Nordström BH to leading order in small
frequencies (which turns out to be a crucial ingredient of our near-extremal
analysis). Furthermore, we explicitly obtain the near-extremal Hawking-evaporation
rate ($\propto\Delta^{4}$), with an analytical expression for the
prefactor (obtained here for the first time to the best of our knowledge).
\end{abstract}
\maketitle

\section{Introduction.}

This paper extends our previous one \citep{FluxesIH:2020}, in which
we computed and investigated the semiclassical stress-energy fluxes
at the inner horizon (IH) of a spherical charged black hole (BH).
Whereas in the previous paper we considered BHs with a broad range
of $Q/M$ values, here we restrict our attention to the near-extremal
limit where $Q/M$ approaches unity, where $Q$ and $M$ respectively
denote the BH's charge and mass.

The \emph{semiclassical} formulation of general relativity treats
matter fields as quantum fields, propagating on a spacetime background
described by a classical metric $g_{\alpha\beta}(x^{\mu})$. The classical
Einstein field equation is then replaced by its semiclassical counterpart:

\[
G_{\alpha\beta}=8\pi\left\langle T_{\alpha\beta}\right\rangle _{\text{ren}}
\]
where $G_{\alpha\beta}$ is the Einstein tensor associated with $g_{\alpha\beta}$
and $\left\langle T_{\alpha\beta}\right\rangle _{\text{ren}}$ is
the renormalized expectation value of the stress-energy tensor (RSET)
associated with the quantum field in consideration. Evidently, the
quantum field and the geometry of spacetime undergo mutual influence.
In particular, the curved geometry of spacetime induces a non-trivial
stress-energy tensor, even in vacuum states, which in turn deforms
the underlying background geometry --- a phenomenon known as \emph{backreaction}.

As our spacetime background, we hereby consider a spherical charged
BH given in the standard Schwarzschild coordinates by the Reissner-Nordström
(RN) metric,

\[
\text{d}s^{2}=-f\left(r\right)\text{d}t^{2}+f^{-1}\left(r\right)\text{d}r^{2}+r^{2}\left(\text{d}\theta^{2}+\sin^{2}\theta\text{d}\phi^{2}\right)\,,
\]
with the $r$-dependent function $f\left(r\right)=1-\frac{2M}{r}+\frac{Q^{2}}{r^{2}}$.
We consider a non-extremal RN BH, meaning $0<Q/M<1$ \footnote{Since the metric doesn't depend on the sign of $Q$, we take without
loss of generality $Q>0$.}. This BH metric admits two horizons, corresponding to the two real
roots of $f\left(r\right)$, denoted by $r_{\pm}=M\pm\sqrt{M^{2}-Q^{2}}$;
the event horizon (EH) is located at $r=r_{+}$, while the IH is located
at $r=r_{-}$. For later use, we also define the two surface gravity
parameters, $\kappa_{\pm}=\left(r_{+}-r_{-}\right)/2r_{\pm}^{2}$.

Upon this BH background we introduce an (uncharged) minimally-coupled
massless scalar field $\Phi$, evolving according to the Klein-Gordon
equation
\[
\square\Phi=0\,,
\]
with $\square$ the covariant d'Alembertian associated with the RN
metric. This field may be decomposed into standard $\omega lm$ modes,
which, due to the RN metric symmetries, can be factorized into a $t$-dependent
term $e^{-i\omega t}$, an angular term $Y_{lm}\left(\theta,\phi\right)$,
and a term that depends on $r$ alone. We cast this last term as $\psi_{\omega l}\left(r\right)/r$,
where $\psi_{\omega l}\left(r\right)$ is the so-called \emph{radial
function} obeying the \emph{radial equation}:
\begin{equation}
\frac{\text{d}^{2}\psi_{\omega l}\left(r\right)}{\text{d}r_{*}^{2}}=\left[V_{l}\left(r\right)-\omega^{2}\right]\psi_{lm}\left(r\right)\,,\label{eq:rad_eq}
\end{equation}
where the $r$-dependent effective potential $V_{l}\left(r\right)$
is given by
\begin{equation}
V_{l}\left(r\right)=f\left(r\right)\left[\frac{l\left(l+1\right)}{r^{2}}+\frac{\text{d}f/\text{d}r}{r}\right]\,,\label{eq:eff_pot}
\end{equation}
and $r_{*}$ is the tortoise coordinate defined through $\text{d}r/\text{d}r_{*}=f\left(r\right)$.
\footnote{Note that there is a freedom of an additive integration constant in
the definition of $r_{*}$, but the analysis which follows is independent
of such a choice.}

We shall consider our field in two vacuum quantum states: the \emph{Hartle-Hawking}
(HH) state \citep{HH:1976,Israel:1976}, corresponding to a quantum
field in thermal equilibrium with an infinite bath of radiation, and
the more physically feasible \emph{Unruh} state \citep{Unruh:1976},
describing the quantum state of a BH that evaporates via Hawking radiation.

We introduce the future-directed null Eddington coordinates, given
inside the BH by $u=r_{*}-t$ and $v=r_{*}+t$. The $\left\langle T_{uu}\right\rangle _{\text{ren}}$
and $\left\langle T_{vv}\right\rangle _{\text{ren}}$ components of
the RSET are referred to as the \emph{flux components}, as they may,
for example, describe correspondingly an ingoing and an outgoing flux
of radiation. In the HH state, time-inversion symmetry implies $\left\langle T_{uu}\left(r\right)\right\rangle _{\text{ren}}=\left\langle T_{vv}\left(r\right)\right\rangle _{\text{ren}}$.
In both quantum states, energy-momentum conservation yields the constancy
(namely $r$-independence) of the quantity 
\begin{equation}
4\pi r^{2}\left\langle T_{t}^{r}\right\rangle _{\text{ren}}=4\pi r^{2}\left(\left\langle T_{uu}\right\rangle _{\text{ren}}-\left\langle T_{vv}\right\rangle _{\text{ren}}\right)\,.\label{eq:conserved}
\end{equation}
In the HH state this constant trivially vanishes. In the Unruh state,
it coincides with the Hawking outflux (as may be seen by evaluating
the above expression at $r\to\infty$, noting that in the Unruh state
$r^{2}\left\langle T_{vv}\right\rangle _{\text{ren}}$ vanishes in
that limit).

As discussed in Ref. \citep{FluxesIH:2020}, the flux components are
crucial for backreaction in the vicinity of the IH, potentially having
an accumulating effect on the form of the metric there. We thus concentrate
on the IH value of the three flux quantities, $\left\langle T_{uu}^{-}\right\rangle _{\text{ren}}^{H}$,
$\left\langle T_{uu}^{-}\right\rangle _{\text{ren}}^{U}$ and $\left\langle T_{vv}^{-}\right\rangle _{\text{ren}}^{U}$,
where the superscript ''$H$'' (''$U$'') corresponds to the HH
(Unruh) state, and an upper ``$-$'' indicates the IH limit. (Hereafter,
the term \emph{flux quantities} will refer to these three IH quantities.)
In Ref. \citep{FluxesIH:2020}, we computed the near-IH flux components
in both quantum states for a variety of non-extremal RN BHs, and displayed
the results as a function of $Q/M$ (for a related work, see also
Ref. \citep{Hollands:2020}). All three flux quantities, $\left\langle T_{uu}^{-}\right\rangle _{\text{ren}}^{H}$,
$\left\langle T_{uu}^{-}\right\rangle _{\text{ren}}^{U}$ and $\left\langle T_{vv}^{-}\right\rangle _{\text{ren}}^{U}$,
were found to change their sign at some $Q/M$ value (all around $\approx0.967$),
being increasingly positive at lower $Q/M$ values and becoming negative
beyond that critical $Q/M$ value. Furthermore, as $Q/M$ grows towards
the extremal value of $1$, all flux quantities decay to zero (at
different rates).

Here, we intend to take a closer look at the near-extremal limit,
characterized by $1-Q/M\ll1$. That is, we wish to examine the near-IH
fluxes as $Q/M$ approaches 1. As we shall see, the near-extremal
domain lends itself to analytical investigation, which sheds light
on the behavior we see numerically. In fact, we find it beneficial
to focus on an equivalent set of three quantities, being the elementary
flux quantity $\left\langle T_{uu}^{-}\right\rangle _{\text{ren}}^{H}$
and the differences $\left\langle T_{uu}^{-}\right\rangle _{\text{ren}}^{H}-\left\langle T_{uu}^{-}\right\rangle _{\text{ren}}^{U}$
and $\left\langle T_{uu}^{-}\right\rangle _{\text{ren}}^{U}-\left\langle T_{vv}^{-}\right\rangle _{\text{ren}}^{U}$
\footnote{Clearly, this set is equivalent (in the sense of the encoded information)
to the basic triplet of flux quantities, with $\left\langle T_{uu}^{-}\right\rangle _{\text{ren}}^{H}$
being the anchoring quantity shared by the two sets.}. The study of the differences, rather than the flux quantities directly,
allows a sharper investigation of the near-extremal domain, as these
differences vanish faster than their constituents on approaching extremality.

One obvious motivation to consider the near-extremal limit is the
very evaporation process considered here: Since our scalar field $\Phi$
is uncharged, the BH charge remains fixed at all times, while the
mass steadily shrinks due to the emission of Hawking radiation. In
the long run, the BH mass $M$ will decay asymptotically to $Q$.
As $M$ approaches $Q$, the Hawking temperature vanishes and the
evaporation rate decays to zero. Note that in such an evaporation
process the BH lasts forever, approaching extremality at the late-time
limit (for a detailed discussion of this evaporation process, see
Ref. \citep{Jacobson}).\footnote{We should bear in mind, however, that this scenario is not particularly
realistic, due to the abundance of charged particles (e.g. in the
form of plasma) in the universe, efficiently acting to neutralize
charged BHs.}

To compute the quantum fluxes in the IH-vicinity, we shall employ
the $\theta$-splitting variant \citep{AAtheta:2016,LeviThetaRSET}
of the ``pragmatic mode-sum regularization'' (PMR) method \citep{AAt:2015,AARSET:2016,LeviRSET:2017},
as we did in Ref. \citep{FluxesIH:2020}. Here, however, owing to
the notable simplicity of the near-extremal limit, we shall carry
this computation mostly analytically, and then validate our analytical
results against numerical ones.

The rest of the paper is organized as follows. In Sec. \ref{sec:Preliminaries.}
we develop the required preliminaries for the near-extremal analysis.
Sec. \ref{sec:The-near-extremal-fluxes} serves as the core of the
paper, in which we perform the analysis of the flux quantities and
their differences in the near-extremal limit. Numerical results and
their agreement with the expressions found in the previous section
are presented in Sec. \ref{sec:Numerical-results.}. We end with a
summary of our results and a short discussion in Sec. \ref{sec:Discussion.}.
In the Appendix we analyze the transmission coefficient outside the
BH to leading order in low frequencies, a quantity required for our
analysis.

In this paper, we work in general-relativistic units $c=G=1$ and
signature $(-+++)$. 

\section{Preliminaries.\label{sec:Preliminaries.}}

In this section we lay the foundations for the near-extremal analysis.
The first subsection presents the basic expressions for the three
flux quantities at the IH, as developed in Ref. \citep{FluxesIH:2020},
from which we construct the three quantities to be focused on in this
paper. The second subsection is devoted to analyzing the internal
radial function in the near-extremal limit, particularly in the vicinity
of the IH.

\subsection{Basic expressions for the fluxes and their differences at the IH.\label{subsec:Basic-expressions-nearIH}}

In the BH interior, we endow the radial equation (\ref{eq:rad_eq})
with the initial condition of a free incoming wave at the EH \footnote{In the BH interior $r$ is timelike, and so is $r_{*}$. $r$ is monotonically
decreasing with time, whereas $r_{*}$ is monotonically increasing.
The EH ($r=r_{+}$) is in fact the past boundary of the BH interior,
and it corresponds to $r_{*}\to-\infty$.}: 
\begin{equation}
\psi_{\omega l}\cong e^{-i\omega r_{*}}\,,\,\,\,\,\,\,r_{*}\to-\infty\,.\label{eq:psi_at_EH}
\end{equation}
At the other edge, in the IH vicinity, the effective potential (\ref{eq:eff_pot})
vanishes like $r-r_{-}$ . Hence, the radial function (satisfying
Eq. (\ref{eq:rad_eq})) attains the general free asymptotic form:
\begin{equation}
\psi_{\omega l}\cong A_{\omega l}e^{i\omega r_{*}}+B_{\omega l}e^{-i\omega r_{*}}\,,\,\,\,\,\,\,r_{*}\to\infty\label{eq:AandB}
\end{equation}
with $A_{\omega l}$ and $B_{\omega l}$ some constant coefficients
determined by the scattering inside the BH. Notably, $A_{\omega l}$
and $B_{\omega l}$ satisfy the relation
\begin{equation}
\left|B_{\omega l}\right|^{2}-\left|A_{\omega l}\right|^{2}=1\,,\label{eq:wronskian}
\end{equation}
arising from the invariance of the Wronskian of $\psi_{\omega l}$
and its complex conjugate.

The basic quantities we concentrate on hereafter involve $A_{\omega l}$
and $B_{\omega l}$, as well as the transmission and reflection coefficients
$\tau_{\omega l}^{\text{up}}$ and $\rho_{\omega l}^{\text{up}}$
for the ``up'' modes scattered outside the BH (see definition in
Ref. \citep{Group:2018}). We shall analyze the near-extremal limit
of $A_{\omega l}$ and $B_{\omega l}$ in Sec. \ref{subsec:Near-extremal-internal-scattering},
while an analysis of $\tau_{\omega l}^{\text{up}}$ and $\rho_{\omega l}^{\text{up}}$
is deferred to the Appendix.

As mentioned in the introduction, we shall be interested in the flux
components $\left\langle T_{uu}\right\rangle _{\text{ren}}$ and $\left\langle T_{vv}\right\rangle _{\text{ren}}$
in both quantum states, in the vicinity of the IH. In Ref. \citep{FluxesIH:2020}
we obtained expressions for these three elementary flux quantities,
$\left\langle T_{uu}^{-}\right\rangle _{\text{ren}}^{H}$, $\left\langle T_{uu}^{-}\right\rangle _{\text{ren}}^{U}$
and $\left\langle T_{vv}^{-}\right\rangle _{\text{ren}}^{U}$, as
a regularized mode sum, employing the $\theta$-splitting variant
of the PMR method. We hereby quote the resulting expressions for convenience
(see Eqs. (9)-(13) therein).

The flux quantities at the IH are generally given by

\begin{align}
\left\langle T_{yy}^{-}\right\rangle _{\text{ren}}^{\Xi} & =\hbar\sum_{l=0}^{\infty}\frac{2l+1}{8\pi}\left(F_{l\left(yy\right)}^{\Xi}-\beta\right)\,,\label{eq: Final_H}
\end{align}
where the superscript ``$\Xi$`` denotes the state (either $H$
or $U$), the subscript ``$y$`` stands for either $u$ or $v$,
\[
F_{l\left(yy\right)}^{\Xi}\equiv\int_{0}^{\infty}d\omega\,\hat{E}_{\omega l\left(yy\right)}^{\Xi}\,,
\]
and $\beta$ is a constant (the so-called ``blind-spot''; to be
given explicitly in Eq. (\ref{eq:plateau}) below), which is the large-$l$
limit of $F_{l\left(yy\right)}^{\Xi}$. The integrand $\hat{E}_{\omega l\left(yy\right)}$
for the HH state is:
\begin{equation}
\hat{E}_{\omega l\left(yy\right)}^{H}=\frac{\omega}{\pi r_{-}^{2}}\left[\coth\left(\pi\omega/\kappa_{+}\right)\left|A_{\omega l}\right|^{2}+\text{csch}\left(\pi\omega/\kappa_{+}\right)\,\Re\left(\rho_{\omega l}^{\text{up}}A_{\omega l}B_{\omega l}\right)\right]\,\label{eq:HH_integrand}
\end{equation}
(where $\Re$ denotes the real part and $\text{csch}\equiv1/\sinh$),
and the corresponding integrands in the Unruh state are given by:
\begin{align}
\hat{E}_{\omega l\left(uu\right)}^{U} & =\hat{E}_{\omega l\left(yy\right)}^{H}+\frac{\omega}{2\pi r_{-}^{2}}\left[1-\coth\left(\pi\omega/\kappa_{+}\right)\right]\left|\tau_{\omega l}^{\text{up}}\right|^{2}\left|A_{\omega l}\right|^{2}\,,\label{eq:U_uu_integrand}
\end{align}
\begin{align}
\hat{E}_{\omega l\left(vv\right)}^{U} & =\hat{E}_{\omega l\left(yy\right)}^{H}+\frac{\omega}{2\pi r_{-}^{2}}\left[1-\coth\left(\pi\omega/\kappa_{+}\right)\right]\left|\tau_{\omega l}^{\text{up}}\right|^{2}\left|B_{\omega l}\right|^{2}\label{eq:U_vv_integrand}
\end{align}

Note that the difference between any of the two Unruh integrands and
the HH integrand goes like $\propto\left|\tau_{\omega l}^{\text{up}}\right|^{2}$,
which in turn decays with $l$ for fixed $\omega$ (note that the
potential barrier outside the BH, given in Eq. (\ref{eq:eff_pot}),
goes like $l\left(l+1\right)$, thus blocking the transmission at
large $l$). Hence, all three flux quantities share the same large-$l$
``blind-spot'' $\beta$, which may be analytically derived (see
Sec. 3 in the Supplemental Material of Ref. \citep{FluxesIH:2020})
to be given by:
\begin{equation}
\beta=\frac{1}{24\pi r_{-}^{2}}\left(\kappa_{-}^{2}-\kappa_{+}^{2}\right)\,.\label{eq:plateau}
\end{equation}

The three flux quantities $\left\langle T_{uu}^{-}\right\rangle _{\text{ren}}^{H}$,
$\left\langle T_{uu}^{-}\right\rangle _{\text{ren}}^{U}$ and $\left\langle T_{vv}^{-}\right\rangle _{\text{ren}}^{U}$
(to which we shall hereafter also refer collectively as the \emph{elementary
triplet}) were the focus of our previous paper \citep{FluxesIH:2020},
where they were computed for a wide variety of subextremal $Q/M$
values. However, in the near-extremal domain, we find it worthwhile
to organize these three flux quantities in a different manner. That
is, we shall focus on an equivalent, slightly different, set of three
quantities (to which we shall occasionally refer as the \emph{derived
triplet}): (\emph{i}) the near-IH flux component in the HH state,
$\left\langle T_{uu}^{-}\right\rangle _{\text{ren}}^{H}$, which also
equals $\left\langle T_{vv}^{-}\right\rangle _{\text{ren}}^{H}$;
(\emph{ii}) the difference between the HH and Unruh values of $\left\langle T_{uu}^{-}\right\rangle _{\text{ren}}$,
which we shall denote by $\left\langle T_{uu}^{-}\right\rangle _{\text{ren}}^{H-U}\equiv\left\langle T_{uu}^{-}\right\rangle _{\text{ren}}^{H}-\left\langle T_{uu}^{-}\right\rangle _{\text{ren}}^{U}$;
and (\emph{iii}) the difference between the two near-IH flux components
in the Unruh state, multiplied by $4\pi r_{-}^{2}$, namely $\Lambda\equiv4\pi r_{-}^{2}\left(\left\langle T_{uu}^{-}\right\rangle _{\text{ren}}^{U}-\left\langle T_{vv}^{-}\right\rangle _{\text{ren}}^{U}\right)$.
Other than its interesting behavior in the near-extremal domain, considering
$\Lambda$ has further motivation -- one may recognize it as the
conserved quantity mentioned in Eq. (\ref{eq:conserved}), in the
Unruh state, evaluated at the IH. \footnote{Hence, we shall hereafter often refer to $\Lambda$ as the ``conserved
quantity''.} Obviously, since this quantity is independent of $r$, its value
may also be evaluated outside the BH. In this sense, $\Lambda$ is
the simplest quantity of all three members of the derived triplet,
as it is fully determined by the scattering problem outside the BH.

The first quantity, $\left\langle T_{uu}^{-}\right\rangle _{\text{ren}}^{H}$,
is given in Eqs. (\ref{eq: Final_H}), (\ref{eq:HH_integrand}) and
(\ref{eq:plateau}). The second quantity $\left\langle T_{uu}^{-}\right\rangle _{\text{ren}}^{H-U}$
is determined from Eqs. (\ref{eq: Final_H}) and (\ref{eq:U_uu_integrand}),
or explicitly:

\begin{equation}
\left\langle T_{uu}^{-}\right\rangle _{\text{ren}}^{H-U}=\hbar\sum_{l=0}^{\infty}\frac{2l+1}{4\pi}\int_{0}^{\infty}d\omega\,\frac{\omega}{4\pi r_{-}^{2}}\left[\coth\left(\pi\omega/\kappa_{+}\right)-1\right]\left|\tau_{\omega l}^{\text{up}}\right|^{2}\left|A_{\omega l}\right|^{2}\,.\label{eq:dTuu}
\end{equation}
Finally, the third quantity $\Lambda$ is obtained by subtracting
Eq. (\ref{eq:U_uu_integrand}) from Eq. (\ref{eq:U_vv_integrand})
and using the Wronskian relation (\ref{eq:wronskian}):

\begin{align}
\Lambda=\hbar\sum_{l=0}^{\infty}\frac{2l+1}{4\pi}\int_{0}^{\infty}d\omega\,\omega\left[\coth\left(\pi\omega/\kappa_{+}\right)-1\right]\left|\tau_{\omega l}^{\text{up}}\right|^{2} & \,.\label{eq:Lambda}
\end{align}
As expected, this conserved quantity only requires the transmission
coefficient outside the BH. Indeed, this is the known expression for
the luminosity of an evaporating BH (see, for example, Eq. (136) in
Ref. \citep{DeWitt:1975} or Eq. (6.20) in Ref. \citep{ChristensenFulling:1977}
for the Schwarzschild case. The only modification needed is replacing
the Schwarzschild $\kappa$ parameter by the corresponding RN parameter
$\kappa_{+}$). 

In Sec. \ref{sec:The-near-extremal-fluxes} we shall take the above
expressions for the derived triplet of quantities, which are valid
for any $Q/M$, and evaluate them in the near-extremal domain of $Q/M$
approaching $1$.

\subsection{The rescaled radial equation.}

To quantify near-extremality, we define the dimensionless parameter
$\Delta$ to be half the difference between $r_{+}/M$ and $r_{-}/M$:
\begin{equation}
\Delta\equiv\sqrt{1-\left(Q/M\right)^{2}}=r_{+}/M-1=1-r_{-}/M\,.\label{eq:Delta}
\end{equation}
Note that $\Delta$ varies from $1$ (Schwarzschild) to $0$ (extremal
RN), whereas the near-extremal domain is characterized by $\Delta\ll1$.
We shall examine the behavior of the various quantities upon approaching
extremality by constructing their leading-order expansions in small
$\Delta$.

To analyze the scaling with $\Delta$, it may be helpful to rewrite
the radial equation (\ref{eq:rad_eq}) in a $\Delta$-normalized fashion,
as we shall now demonstrate. 

In the BH interior, the radial variable $r$ is confined to a domain
of width $2M\Delta$,
\[
1-\Delta\leq r/M\leq1+\Delta\,.
\]
That is, $r/M-1$ scales linearly with $\Delta$. We thus define the
rescaled variable
\[
s\equiv\frac{r/M-1}{\Delta}\,,
\]
suitable for our near-extremal analysis. Note that $s$ varies from
$1$ at the EH to $-1$ at the IH. One finds that the function $f\left(r\right)$
is:
\[
f=\Delta^{2}\frac{s^{2}-1}{\left(1+\Delta\,s\right)^{2}}\,,
\]
and the effective potential $V_{l}$ (\ref{eq:eff_pot}) written in
terms of the variable $s$ is:
\[
V_{l}=\frac{\Delta^{2}}{M^{2}}\frac{s^{2}-1}{\left(1+\Delta\,s\right)^{4}}\left[l\left(l+1\right)+2\Delta\frac{s+\Delta}{\left(1+\Delta\,s\right)^{2}}\right]\,.
\]

We now write this effective potential separately for $l=0$ and $l>0$,
expressed in each of these two cases at its leading order in the small
parameter $\Delta$:
\begin{equation}
V_{l=0}=2\frac{\Delta^{3}}{M^{2}}\,s\left(s^{2}-1\right)+\mathcal{O}\left(\Delta^{4}\right)\label{eq:V_l=00003D0}
\end{equation}
for $l=0$, and
\begin{equation}
V_{l>0}=\frac{\Delta^{2}}{M^{2}}l\left(l+1\right)\left(s^{2}-1\right)+\mathcal{O}\left(\Delta^{3}\right)\,\label{eq:V_l>0}
\end{equation}
for $l>0$.

The variable $s$ is related to $r_{*}$ via
\[
\frac{\text{d}s}{\text{d}r_{*}}=\frac{f}{M\Delta}=\frac{\Delta}{M}\left(s^{2}-1\right)+\mathcal{O}\left(\Delta^{2}\right)\,,
\]
meaning that $r_{*}$ basically scales like $M/\Delta$. We thus define
the rescaled dimensionless variable $\tilde{r}_{*}\equiv(\Delta/M)r_{*}$.
It satisfies the ODE 
\[
\frac{\text{d}s}{\text{d}\tilde{r}_{*}}=\left(s^{2}-1\right)+\mathcal{O}\left(\Delta\right)\,,
\]
which may be solved to yield
\begin{equation}
s\left(\tilde{r}_{*}\right)=-\tanh\left(\tilde{r}_{*}\right)+\mathcal{O}\left(\Delta\right)\,.\label{eq:  s(r*)}
\end{equation}

We also define the rescaled dimensionless frequency and effective
potential, $\tilde{\omega}\equiv(M/\Delta)\omega$ and $\tilde{V}_{l}\equiv(M^{2}/\Delta^{2})V_{l}$,
respectively. We may now rewrite the radial equation (\ref{eq:rad_eq})
in a rescaled fashion, in the variable $\tilde{r}_{*}$, as

\begin{equation}
\psi_{\omega l,\tilde{r}_{*}\tilde{r}_{*}}=\left(\tilde{V}_{l}-\tilde{\omega}^{2}\right)\psi_{\omega l}\,,\label{eq:rescaled_radial_equation}
\end{equation}
along with the boundary condition $\psi_{\omega l}\cong e^{-i\tilde{\omega}\tilde{r}_{*}}$
at $\tilde{r}_{*}\to-\infty$ (in correspondence with Eq. (\ref{eq:psi_at_EH})). 

Finally, we use Eq. (\ref{eq:  s(r*)}) to rewrite the rescaled potentials
for $l=0$ (\ref{eq:V_l=00003D0}) and $l>0$ (\ref{eq:V_l>0}) explicitly
in terms of $\tilde{r}_{*}$, to leading order in $\Delta$:
\begin{equation}
\tilde{V}_{l=0}=2\Delta\tanh\left(\tilde{r}_{*}\right)\text{sech}^{2}\left(\tilde{r}_{*}\right)+\mathcal{O}\left(\Delta^{2}\right)\label{eq:Vt_l=00003D0}
\end{equation}
and
\begin{equation}
\tilde{V}_{l>0}=-l\left(l+1\right)\text{sech}^{2}\left(\tilde{r}_{*}\right)+\mathcal{O}\left(\Delta\right)\,.\label{eq:Vt_l>0}
\end{equation}

\subsubsection{Near-extremal internal scattering.\label{subsec:Near-extremal-internal-scattering}}

We are interested in the $\Delta\ll1$ limit of the coefficients $A_{\omega l}$
and $B_{\omega l}$ appearing in the near-IH free asymptotic form
of the radial function (\ref{eq:AandB}), as they are vital components
in the quantities we wish to analyze (as seen in Subsec. \ref{subsec:Basic-expressions-nearIH}).
The rescaled radial equation (\ref{eq:rescaled_radial_equation})
developed above may be analyzed to solve the scattering problem in
the BH interior to leading order in $\Delta$, yielding $A_{\omega l}$
and $B_{\omega l}$ to that order.

\paragraph{The $l=0$ case}

For $l=0$, the rescaled potential (\ref{eq:Vt_l=00003D0}) vanishes
like $\Delta$. That is, in the near-extremal domain $\tilde{V}_{l=0}\ll\tilde{\omega}^{2}$
(for any given $\tilde{\omega}>0$), hence the radial equation for
$l=0$ lends itself to a leading-order Born approximation. Accordingly,
the asymptotic behavior of the radial function at $\tilde{r}_{*}\to\infty$
is:
\begin{align}
\psi_{\omega,l=0} & \cong e^{-i\tilde{\omega}\tilde{r}_{*}}\left(1-\frac{1}{2i\tilde{\omega}}\int_{-\infty}^{\infty}\tilde{V}_{l=0}\left(x\right)dx\right)+\frac{e^{i\tilde{\omega}\tilde{r}_{*}}}{2i\tilde{\omega}}\int_{-\infty}^{\infty}e^{-2i\tilde{\omega}x}\tilde{V}_{l=0}\left(x\right)dx\nonumber \\
 & =e^{-i\tilde{\omega}\tilde{r}_{*}}-2\pi\Delta\tilde{\omega}\text{csch}\left(\pi\tilde{\omega}\right)e^{i\tilde{\omega}\tilde{r}_{*}}+\mathcal{O}\left(\Delta^{2}\right)\,.\label{eq:psi_l=00003D0_interior}
\end{align}
Note that the term $\int_{-\infty}^{\infty}\tilde{V}_{l=0}\left(x\right)dx$
leaves $\mathcal{O}\left(\Delta^{2}\right)$, owing to the odd parity
of the leading order of $\tilde{V}_{l=0}$ (see Eq. (\ref{eq:Vt_l=00003D0})).

Comparing this with the asymptotic form (\ref{eq:AandB}) we get the
coefficients $A_{\omega l}$ and $B_{\omega l}$ at $l=0$, to leading
order in $\Delta$, to be:
\begin{align}
A_{\omega,l=0}=-2\pi\Delta\tilde{\omega}\text{csch}\left(\pi\tilde{\omega}\right)+\mathcal{O}\left(\Delta^{2}\right)\,\,\,,\,\,\,B_{\omega,l=0}=1+\mathcal{O}\left(\Delta^{2}\right)\,.\label{eq:AandB_at_l=00003D0}
\end{align}

\paragraph{The $l>0$ case}

Note that unlike the $l=0$ case, for $l>0$ Eq. (\ref{eq:rescaled_radial_equation})
with the rescaled potential (\ref{eq:Vt_l>0}) is insensitive to $\Delta$.
The scattering problem is given (to leading order in $\Delta$) by
the corresponding equation:
\[
\psi_{\omega l,\tilde{r}_{*}\tilde{r}_{*}}=-\left[l\left(l+1\right)\text{sech}^{2}\left(\tilde{r}_{*}\right)+\tilde{\omega}^{2}\right]\psi_{\omega l}\,,
\]
and it is solved analytically to yield 
\[
\psi_{\omega,l>0}=c_{1}P_{l}^{i\tilde{\omega}}\left(z\right)+c_{2}Q_{l}^{i\tilde{\omega}}\left(z\right)
\]
where $P_{l}^{i\tilde{\omega}}$ is the associated Legendre polynomial,
$Q_{l}^{i\tilde{\omega}}$ is the associated Legendre function of
the second kind, $c_{1}$ and $c_{2}$ are coefficients to be determined,
and we define the variable $z\equiv-\tanh\tilde{r}_{*}$. \footnote{$z$ actually coincides with $s$ to leading order in $\Delta$, see
Eq. (\ref{eq:  s(r*)}).} Note that $z\to1$ corresponds to the EH, whereas $z\to-1$ corresponds
to the IH.

In order to find $c_{1}$ and $c_{2}$, we carry the above general
solution to the EH, noting that
\[
P_{l}^{i\tilde{\omega}}\left(z\to1\right)\cong\frac{1}{\Gamma\left(1-i\tilde{\omega}\right)}\left(\frac{1-z}{2}\right)^{-i\tilde{\omega}/2}\,,
\]
where $\Gamma$ hereafter denotes the gamma function, as well as
\[
Q_{l}^{i\tilde{\omega}}\left(z\to1\right)\cong\left(\frac{1-z}{2}\right)^{-i\tilde{\omega}/2}\frac{\cosh\left(\pi\tilde{\omega}\right)\Gamma\left(i\tilde{\omega}\right)}{2}-\left(\frac{1-z}{2}\right)^{i\tilde{\omega}/2}\frac{\Gamma\left(-l+i\tilde{\omega}\right)\Gamma\left(-i\tilde{\omega}\right)}{2\Gamma\left(-l-i\tilde{\omega}\right)}\,.
\]
 In addition, note that at $z\to1$ we have $1-z\cong2e^{2\tilde{r}_{*}}$,
and thus
\[
\left(\frac{1-z}{2}\right)^{-i\tilde{\omega}/2}\cong e^{-i\tilde{\omega}\tilde{r}_{*}}=e^{-i\omega r_{*}}\,.
\]
 Then, matching with the initial condition (\ref{eq:psi_at_EH}) of
a free incoming wave at the EH yields $c_{1}=\Gamma\left(1-i\tilde{\omega}\right)$
along with $c_{2}=0$. That is, the radial function for $l>0$ in
the BH interior is
\begin{equation}
\psi_{\omega,l>0}=\Gamma\left(1-i\tilde{\omega}\right)P_{l}^{i\tilde{\omega}}\left(z\right)\,.\label{eq:psi_l>0_interior}
\end{equation}

Now, in order to carry the above expression to the IH, we note that
\[
P_{l}^{i\tilde{\omega}}\left(z\to-1\right)\cong i\left(\frac{1+z}{2}\right)^{i\tilde{\omega}/2}\frac{\pi\text{csch}\left(\pi\tilde{\omega}\right)}{\Gamma\left(-l-i\tilde{\omega}\right)\Gamma\left(1+l-i\tilde{\omega}\right)\Gamma\left(1+i\tilde{\omega}\right)}
\]
and that at $z\to-1$, following $1+z\cong2e^{-2\tilde{r}_{*}}$,
\[
\left(\frac{1+z}{2}\right)^{i\tilde{\omega}/2}\cong e^{-i\tilde{\omega}\tilde{r}_{*}}=e^{-i\omega r_{*}}\,.
\]
Then, comparing with the free asymptotic form in Eq. (\ref{eq:AandB})
yields:

\begin{equation}
A_{\omega,l>0}=\mathcal{O}\left(\Delta\right),\,\,\,B_{\omega,l>0}=i\frac{\pi\,\text{csch}\left(\pi\tilde{\omega}\right)\Gamma\left(1-i\tilde{\omega}\right)}{\Gamma\left(-l-i\tilde{\omega}\right)\Gamma\left(1+l-i\tilde{\omega}\right)\Gamma\left(1+i\tilde{\omega}\right)}+\mathcal{O}\left(\Delta\right)\,.\label{eq:AandB_at_l>0}
\end{equation}

One may verify explicitly that to leading order we have $\left|B_{\omega,l>0}\right|=1$,
as indeed required by Eq. (\ref{eq:wronskian}) given the vanishing
of $A_{\omega,l>0}$. That property, of $A_{\omega l}$ vanishing
with $\Delta$, is actually shared by the $l>0$ and the $l=0$ cases
alike.

\section{The near-extremal flux quantities and their differences.\label{sec:The-near-extremal-fluxes}}

As mentioned previously, in Ref. \citep{FluxesIH:2020} all three
flux quantities $\left\langle T_{uu}^{-}\right\rangle _{\text{ren}}^{H}$,
$\left\langle T_{uu}^{-}\right\rangle _{\text{ren}}^{U}$ and $\left\langle T_{vv}^{-}\right\rangle _{\text{ren}}^{U}$
were computed numerically and shown to decay as $Q/M$ grows towards
$1$ (see Fig. (1) therein). In what follows, we shall focus on the
leading-order behavior in $\Delta$ of the three derived quantities
introduced in Sec. \ref{subsec:Basic-expressions-nearIH}: $\left\langle T_{uu}^{-}\right\rangle _{\text{ren}}^{H}$,
$\left\langle T_{uu}^{-}\right\rangle _{\text{ren}}^{H-U}\equiv\left\langle T_{uu}^{-}\right\rangle _{\text{ren}}^{H}-\left\langle T_{uu}^{-}\right\rangle _{\text{ren}}^{U}$,
and $\Lambda\equiv4\pi r_{-}^{2}\left(\left\langle T_{uu}^{-}\right\rangle _{\text{ren}}^{U}-\left\langle T_{vv}^{-}\right\rangle _{\text{ren}}^{U}\right)$.
The member of this triplet which is simplest to approach is $\Lambda$,
depending on $\tau_{\omega l}^{\text{up}}$ only, as can be seen in
Eq. (\ref{eq:Lambda}). The other two quantities, $\left\langle T_{uu}^{-}\right\rangle _{\text{ren}}^{H}$
and $\left\langle T_{uu}^{-}\right\rangle _{\text{ren}}^{H-U}$, require
both exterior and interior scattering coefficients. We shall, in fact,
treat analytically only two of the three quantities, $\Lambda$ and
$\left\langle T_{uu}^{-}\right\rangle _{\text{ren}}^{H-U}$. The flux
$\left\langle T_{uu}^{-}\right\rangle _{\text{ren}}^{H}$ will not
be treated analytically, but its leading order (based on numerics)
will nevertheless be presented, as a meaningful result. Using the
results for the derived triplet, we subsequently treat the original
three flux quantities $\left\langle T_{yy}^{-}\right\rangle _{\text{ren}}^{\Xi}$.

\subsection{The conserved quantity $\Lambda$ in a near-extremal BH\label{subsec:Lambda_analysis}}

We may evaluate the near-extremal limit of $\Lambda$ through its
mode-sum expression given in Eq. (\ref{eq:Lambda}).

First, note that the surface gravity of a near-extremal RN BH scales
like $\kappa_{+}\cong\Delta/M$. Then, the $\coth\left(\pi\omega/\kappa_{+}\right)-1$
factor in the integrand in Eq. (\ref{eq:Lambda}) (which decays exponentially
in the $\coth$ argument) acts as a weight function on the $\omega$
axis, crucially leaving an effective sampling window of width $\propto\Delta/M$.
We are thus interested in the behavior of the various components of
the integrand in low frequencies. A detailed analysis of $\tau_{\omega l}^{\text{up}}$
to leading order in $\omega M\ll1$ is presented in the Appendix,
with the result given in Eq. (\ref{eq:tau_final}) therein. Notably,
to leading order in low frequencies we have $\tau_{\omega l}^{\text{up}}\propto\omega^{l+1}$
(this holds regardless of $Q/M$, as long as $Q/M<1$). Furthermore,
the prefactor of this leading-order term scales as $\Delta^{l}$.
The contribution of each $l$ to $\Lambda$ (as may be seen through
Eq. (\ref{eq:Lambda})) therefore goes like $\Delta{}^{4\left(l+1\right)}$.
The sum over $l$ in the limit $\Delta\ll1$ is thus dominated by
the $l=0$ term. The transmission coefficient that enters this term
is:
\[
\tau_{\omega,l=0}^{\text{up}}=-2i\omega r_{+}+\mathcal{O}\left(\omega^{2}\right)
\]
(see Eq. (\ref{eq:tau_smallw-l0}) in the Appendix).

We may now proceed to compute $\Lambda$ to leading order in $\Delta$,
using Eq. (\ref{eq:Lambda}) and taking only the $l=0$ contribution
as discussed:
\begin{align}
\Lambda & \cong\hbar\,\frac{r_{+}^{2}}{\pi}\int_{0}^{\infty}d\omega\,\omega^{3}\left[\coth\left(\pi\omega/\kappa_{+}\right)-1\right]\,.\label{eq:Lambda_delta}
\end{align}
Recalling that $\omega=(\Delta/M)\tilde{\omega}\cong\kappa_{+}\tilde{\omega}$
(and also $r_{+}\cong M$), we find it convenient to rewrite this
expression in terms of a dimensionless and $\Delta$-invariant integral: 

\begin{align}
\Lambda & \cong\hbar\,\frac{\Delta^{4}}{\pi M^{2}}\int_{0}^{\infty}d\tilde{\omega}\,\tilde{\omega}^{3}\left[\coth\left(\pi\tilde{\omega}\right)-1\right]=\hbar\,\frac{\Delta^{4}}{120\pi M^{2}}\,.\label{eq:Lambda_delta-2}
\end{align}
We have thus found the Hawking outflux to leading order in $\Delta$
for a near-extremal RN BH. The dependence on $\Delta^{4}$ is well
known (see, e.g., Ref. \citep{Jacobson}), but the prefactor, which
we derived analytically, is given here for the first time as far as
we are aware.

\subsection{$\left\langle T_{uu}^{-}\right\rangle _{\text{ren}}^{H-U}$ in a
near-extremal BH\label{subsec:Tuu^H-U analysis}}

The treatment of $\left\langle T_{uu}^{-}\right\rangle _{\text{ren}}^{H-U}$
closely follows the calculation of $\Lambda$ carried out in the previous
subsection. To this end, it is instructive to compare the expressions
in Eqs. (\ref{eq:dTuu}) and (\ref{eq:Lambda}). Both integrands include
the factor $\coth\left(\pi\omega/\kappa_{+}\right)-1$, implying an
effective frequency window of width $\propto\Delta/M$. In fact, the
only difference between the two integrands (apart from the trivial
constant factor $4\pi r_{-}^{2}$) is the extra multiplicative quantity
$\left|A_{\omega l}\right|^{2}$ appearing in the expression for $\left\langle T_{uu}^{-}\right\rangle _{\text{ren}}^{H-U}$.
As found in Subsec. \ref{subsec:Near-extremal-internal-scattering}
(see Eqs. (\ref{eq:AandB_at_l=00003D0}) and (\ref{eq:AandB_at_l>0})),
the coefficient $A_{\omega l}$ to leading order in $\Delta$ is $\cong-2\pi\Delta\tilde{\omega}\,\text{csch}\left(\pi\tilde{\omega}\right)$
for $l=0$, and vanishes at least like $\Delta$ for $l>0$. We already
established in the previous subsection that the expression (\ref{eq:Lambda})
for $\Lambda$ is dominated by the $l=0$ contribution. Given the
behavior of $A_{\omega l}$ quoted above, the extra $\left|A_{\omega l}\right|^{2}$
factor in the expression for $\left\langle T_{uu}^{-}\right\rangle _{\text{ren}}^{H-U}$
does not alter this situation. Thus, obtaining $\left\langle T_{uu}^{-}\right\rangle _{\text{ren}}^{H-U}$
to leading order in $\Delta$ would merely require multiplying the
integrand in Eq. (\ref{eq:Lambda_delta-2}) by 
\[
\left|A_{\omega,l=0}\right|^{2}\cong[2\pi\Delta\tilde{\omega}\,\text{csch}\left(\pi\tilde{\omega}\right)]^{2}
\]
 (as well as dividing by the constant $4\pi r_{-}^{2}\cong4\pi M^{2}$).
Combining these factors yields:
\begin{align}
\left\langle T_{uu}^{-}\right\rangle _{\text{ren}}^{H-U} & \cong\frac{\hbar}{M^{4}}\Delta^{6}\int_{0}^{\infty}d\tilde{\omega}\,\tilde{\omega}^{5}\left[\coth\left(\pi\tilde{\omega}\right)-1\right]\text{csch}^{2}\left(\pi\tilde{\omega}\right)\nonumber \\
 & =\hbar\frac{\Delta^{6}}{12M^{4}\pi^{6}}\left[\pi^{4}-90\,\zeta\left(5\right)\right]\,,\label{eq:Tuudif_delta}
\end{align}
where $\zeta$ is the Riemann zeta function. 

\subsection{$\left\langle T_{uu}^{-}\right\rangle _{\text{ren}}^{H}$ in a near-extremal
BH\label{subsec:Tuu^H}}

Of the three members of the derived triplet, the expression for $\left\langle T_{uu}^{-}\right\rangle _{\text{ren}}^{H}$
is the most challenging one to analyze. It is given by Eqs. (\ref{eq: Final_H}),
(\ref{eq:HH_integrand}) and (\ref{eq:plateau}). It is fairly easy
to see, for example, that the second term in the integrand in Eq.
(\ref{eq:HH_integrand}) yields a contribution $\propto\Delta^{3}$.
However, there is another such contribution in $\beta$, and these
two $\propto\Delta^{3}$ terms just cancel each other out. Then there
are various potential contributions at order $\propto\Delta^{4}$,
but these are more difficult to analyze, as this analysis would require
computing $A_{\omega l},B_{\omega l}$ and $\rho_{\omega l}^{\text{up}}$
beyond their leading order in $\Delta$. \footnote{Notice that in the analysis of $\Lambda$ and $\left\langle T_{uu}^{-}\right\rangle _{\text{ren}}^{H-U}$,
carried out in the previous two subsections, the corresponding integrands
were both proportional to $\left|\tau_{\omega l}^{\text{up}}\right|^{2}$
(see Eqs. (\ref{eq:dTuu}) and (\ref{eq:Lambda})) along with a weight
factor of effective width $\propto\Delta$ on the $\omega$ axis ---
hence contributing an extra factor $\propto\Delta^{2}$ (and even
higher powers of $\Delta$ for $l>0$). In the present case, no such
$\left|\tau_{\omega l}^{\text{up}}\right|^{2}$ factor exists in the
integrand in Eq. (\ref{eq:HH_integrand}); hence the various potential
contributions start already at lower powers of $\Delta$ compared
to the other two cases.} We therefore resorted here to numerics. A numerical analysis of $\left\langle T_{uu}^{-}\right\rangle _{\text{ren}}^{H}$
indicates that its small-$\Delta$ asymptotic behavior is: 

\begin{align}
\left\langle T_{uu}^{-}\right\rangle _{\text{ren}}^{H} & \cong\alpha\Delta^{5}\,\label{eq:TuuH_delta}
\end{align}
with the numerically extracted coefficient $\alpha\cong-3.4375\times10^{-3}\hbar M^{-4}$.
This behavior is demonstrated in Figs. \ref{Fig:derived_triplet}
and \ref{Fig:elementary_triplet} (by the green dots approaching the
green dashed line). 

\subsection{The three elementary fluxes in a near-extremal BH\label{subsec:elem_triplet}}

In Subsecs. \ref{subsec:Lambda_analysis}, \ref{subsec:Tuu^H-U analysis}
and \ref{subsec:Tuu^H} we analyzed the derived triplet in a near-extremal
RN BH. Here, we shall utilize these results to obtain the leading-order
behavior of the original elementary triplet of fluxes $\left\langle T_{yy}^{-}\right\rangle _{\text{ren}}^{\Xi}$.

Notably, the difference between $\left\langle T_{uu}^{-}\right\rangle _{\text{ren}}^{H}$
and $\left\langle T_{uu}^{-}\right\rangle _{\text{ren}}^{U}$ given
in Eq. (\ref{eq:Tuudif_delta}) decays faster than $\left\langle T_{uu}^{-}\right\rangle _{\text{ren}}^{H}$
(see Eq. (\ref{eq:TuuH_delta})). Consequently, $\left\langle T_{uu}^{-}\right\rangle _{\text{ren}}^{U}$
shares the same leading-order behavior as its HH counterpart, namely
\begin{equation}
\left\langle T_{uu}^{-}\right\rangle _{\text{ren}}^{H}\cong\left\langle T_{uu}^{-}\right\rangle _{\text{ren}}^{U}\cong\alpha\Delta^{5}\label{eq:Tuu_final-1}
\end{equation}
with $\alpha$ as given in the previous subsection.

In addition, recall that $\Lambda$ is proportional to the difference
between $\left\langle T_{uu}^{-}\right\rangle _{\text{ren}}^{U}$
and $\left\langle T_{vv}^{-}\right\rangle _{\text{ren}}^{U}$, and
that it was found to decay like $\Delta^{4}$ (see Eq. (\ref{eq:Lambda_delta-2})).
We thus conclude that in a near-extremal RN BH, the Unruh ingoing
flux component $\left\langle T_{vv}^{-}\right\rangle _{\text{ren}}^{U}$
dominates over its outgoing counterpart $\left\langle T_{uu}^{-}\right\rangle _{\text{ren}}^{U}$,
and approaches $-\Lambda/4\pi r_{-}^{2}\cong-\Lambda/4\pi M^{2}$
as $\Delta$ decreases. Explicitly, the leading order of $\left\langle T_{vv}^{-}\right\rangle _{\text{ren}}^{U}$
in small $\Delta$ is given by:
\begin{equation}
\left\langle T_{vv}^{-}\right\rangle _{\text{ren}}^{U}\cong-\hbar\,\frac{\Delta^{4}}{480\pi^{2}M^{4}}\,.\label{eq:Tvv_delta}
\end{equation}

\section{Numerical results.\label{sec:Numerical-results.}}

Using the methods described in Ref. \citep{FluxesIH:2020}, we computed
the three flux quantities $\left\langle T_{yy}^{-}\right\rangle _{\text{ren}}^{\Xi}$
in a set of $Q/M$ values exponentially approaching the extremal value
of $1$. The procedure includes numerically solving the radial equation
(\ref{eq:rad_eq}) in the BH interior and exterior to extract the
internal scattering coefficients $A_{\omega l}$ and $B_{\omega l}$
(\ref{eq:AandB}) as well as the transmission and reflection coefficients
$\tau_{\omega l}^{\text{up}}$ and $\rho_{\omega l}^{\text{up}}$,
subsequently feeding them into the relevant mode sums as outlined
in Subsec. \ref{subsec:Basic-expressions-nearIH}. Performing the
computation, we found rapid exponential convergence in both $\omega$
and $l$, which facilitates the numerical implementation of the procedure.
\footnote{As may be seen analytically, for each of these three quantities the
integrand decays exponentially with $\omega$ (other than the trivial
decaying factors, this has to do with the analytically-known exponential
decay of $A_{\omega l}$ and $\rho_{\omega l}$ at large frequencies).
The $\omega$ range chosen for the computation suitably scales with
$\Delta$. The series in $l$, constructed after performing the integration
over $\omega$, exhibits too a very quick exponential decay. In fact,
it turns out that at this domain of $\Delta\ll1$ it suffices to include
the $l=0$ contribution alone. Nevertheless, to be on the safe side,
we included a few additional $l$ values in our computation.} Subsequently, from the three flux quantities $\left\langle T_{yy}^{-}\right\rangle _{\text{ren}}^{\Xi}$
we also derived the differences $\left\langle T_{uu}^{-}\right\rangle _{\text{ren}}^{H-U}$
and $\Lambda$.

\begin{figure}[h!]
\centering \includegraphics[width=10cm]{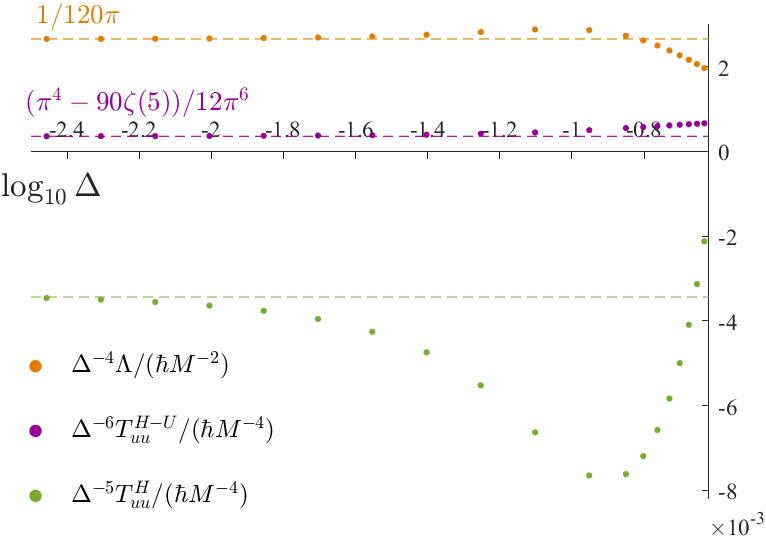}\caption{$\Lambda\Delta^{-4}$, $\left\langle T_{uu}^{-}\right\rangle _{\text{ren}}^{H}\Delta^{-5}$
and $\left\langle T_{uu}^{-}\right\rangle _{\text{ren}}^{H-U}\Delta^{-6}$
in suitable units vs. $\log_{10}\Delta$. The horizontal colored dashed
lines correspond to the coefficients of the leading orders in $\Delta$,
known analytically for $\Lambda$ and $\left\langle T_{uu}^{-}\right\rangle _{\text{ren}}^{H-U}$
as prescribed in Eqs. (\ref{eq:Lambda_delta-2}) and (\ref{eq:Tuudif_delta}).}
\label{Fig:derived_triplet}
\end{figure}

Fig. \ref{Fig:derived_triplet} portrays the leading-order behavior
of the derived triplet $\Lambda$, $\left\langle T_{uu}^{-}\right\rangle _{\text{ren}}^{H}$
and $\left\langle T_{uu}^{-}\right\rangle _{\text{ren}}^{H-U}$, in
the near-extremal domain $\Delta\ll1$. Each flux quantity is divided
by the leading power of $\Delta$ in its near-extremal asymptotic
behavior (namely $\Delta^{4}$, $\Delta^{5}$ and $\Delta^{6}$, respectively).
The approach to extremality amounts to moving leftwards in the figure,
and the figure indicates that all displayed curves flatten at that
limit. The numerical results for $\Lambda$ and $\left\langle T_{uu}^{-}\right\rangle _{\text{ren}}^{H-U}$
are in full agreement with the analytically-derived leading-order
behavior given in Eqs. (\ref{eq:Lambda_delta-2}) and (\ref{eq:Tuudif_delta}),
represented respectively by horizontal orange and purple dashed lines
with the corresponding coefficient values appearing on top. The leading-order
coefficient for $\left\langle T_{uu}^{-}\right\rangle _{\text{ren}}^{H}$
is extracted from the numerics to be $\alpha\simeq-3.4375\times10^{-3}\hbar M^{-4}$,
and is represented by the horizontal green dashed line (in both figures).

\begin{figure}[h!]
\centering \includegraphics[width=10cm]{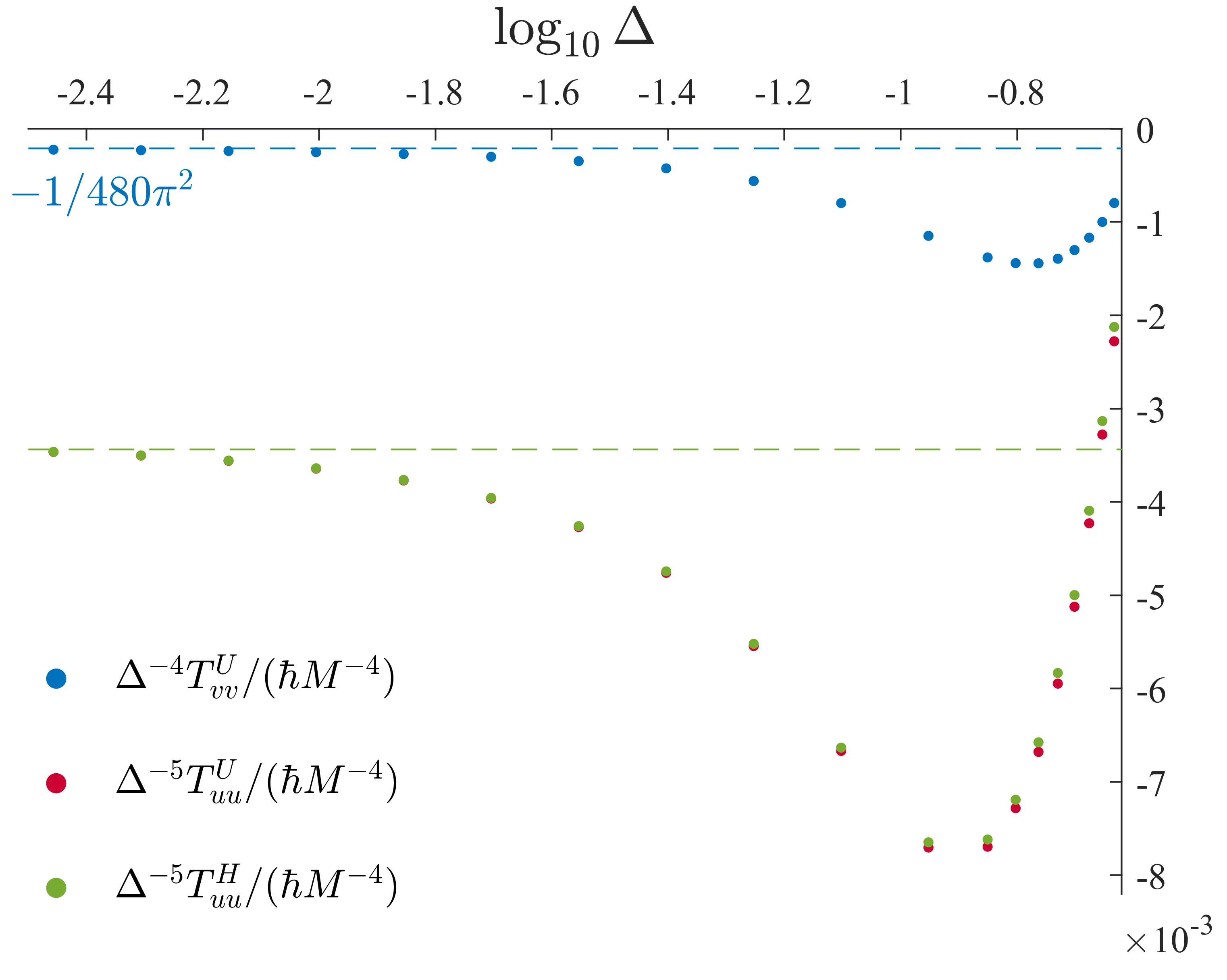}\caption{$\left\langle T_{vv}^{-}\right\rangle _{\text{ren}}^{U}\Delta^{-4}$,
$\left\langle T_{uu}^{-}\right\rangle _{\text{ren}}^{H}\Delta^{-5}$
and $\left\langle T_{uu}^{-}\right\rangle _{\text{ren}}^{U}\Delta^{-5}$
in suitable units vs. $\log_{10}\Delta$. The horizontal colored dashed
lines correspond to the coefficients of the leading orders in $\Delta$,
known to be identical for $\left\langle T_{uu}^{-}\right\rangle _{\text{ren}}^{H}$
and $\left\langle T_{uu}^{-}\right\rangle _{\text{ren}}^{U}$ (see
Eq. \ref{eq:Tuu_final-1}), and known analytically for $\left\langle T_{vv}^{-}\right\rangle _{\text{ren}}^{U}$
as prescribed in Eq. (\ref{eq:Tvv_delta}).}
\label{Fig:elementary_triplet}
\end{figure}

Similarly, Fig. \ref{Fig:elementary_triplet} portrays the leading-order
behavior of the three elementary flux quantities $\left\langle T_{yy}^{-}\right\rangle _{\text{ren}}^{\Xi}$
in the near-extremal domain $\Delta\ll1$. Each flux quantity is divided
by its leading power of $\Delta$ (namely $\Delta^{4}$ or $\Delta^{5}$).
As seen in Eq. (\ref{eq:Tuu_final-1}), $\left\langle T_{uu}^{-}\right\rangle _{\text{ren}}^{H}$
and $\left\langle T_{uu}^{-}\right\rangle _{\text{ren}}^{U}$ share
the same leading order in their expansion in small $\Delta$, hence
their plots coincide towards extremality. The amount by which they
differ has been analyzed and is given in Eq. (\ref{eq:Tuudif_delta})
(and displayed in Fig. \ref{Fig:derived_triplet}). The leading-order
coefficient for $\left\langle T_{vv}^{-}\right\rangle _{\text{ren}}^{U}$
is known analytically (\ref{eq:Tvv_delta}), and is represented by
the blue horizontal dashed line. 

\section{Discussion.\label{sec:Discussion.}}

Our main goal in this paper was to investigate and compute the semiclassical
null fluxes $\left\langle T_{uu}\right\rangle _{\text{ren}}$ and
$\left\langle T_{vv}\right\rangle _{\text{ren}}$ at the IH of a near-extremal
RN BH, in the Unruh and HH quantum states. Since in the HH state we
have $\left\langle T_{vv}\right\rangle _{\text{ren}}=\left\langle T_{uu}\right\rangle _{\text{ren}}$,
there are three such independent flux quantities: $\left\langle T_{uu}^{-}\right\rangle _{\text{ren}}^{U}$,
$\left\langle T_{vv}^{-}\right\rangle _{\text{ren}}^{U}$, and $\left\langle T_{uu}^{-}\right\rangle _{\text{ren}}^{H}$.
(Recall, the ``$-$'' superscript denotes the asymptotic IH value,
and the superscripts ``$U$'' and ``$H$'' respectively refer
to the Unruh and HH quantum states.) We referred to these three flux
quantities as the \emph{elementary triplet} of fluxes. We found it
useful, however, to introduce another (yet mathematically equivalent)
triplet of flux-related quantities: $\left\langle T_{uu}^{-}\right\rangle _{\text{ren}}^{H}$,
$\Lambda$, and $\left\langle T_{uu}^{-}\right\rangle _{\text{ren}}^{H-U}$,
to which we referred as the \emph{derived triplet}. Here ``$H-U$``
denotes the flux difference between the HH and Unruh states, and $\Lambda\equiv4\pi r_{-}^{2}\left(\left\langle T_{uu}^{-}\right\rangle _{\text{ren}}^{U}-\left\langle T_{vv}^{-}\right\rangle _{\text{ren}}^{U}\right)$.
Although the elementary and derived triplets in principle encode the
same information, we found it beneficial to focus our analysis on
the latter triplet, as it allows a sharper investigation of the near-extremal
limit. Firstly, two out of the three members of the derived triplet,
$\Lambda$ and $\left\langle T_{uu}^{-}\right\rangle _{\text{ren}}^{H-U}$,
are amenable to a full leading-order analytical treatment near extremality.
Furthermore, we find that the flux difference $\left\langle T_{uu}^{-}\right\rangle _{\text{ren}}^{H-U}$
decreases faster than both $\left\langle T_{uu}^{-}\right\rangle _{\text{ren}}^{H}$
and $\left\langle T_{uu}^{-}\right\rangle _{\text{ren}}^{U}$ on approaching
extremality. As an additional motivation, $\Lambda$ turns out to
be directly associated with the conserved quantity presented in Eq.
(\ref{eq:conserved}), which in fact coincides with the Hawking-evaporation
outflux to infinity (a point to be further discussed below).

We hereby summarize our findings for the asymptotic behavior of the
various flux quantities, to leading order in the small parameter $\Delta\equiv\sqrt{1-\left(Q/M\right)^{2}}$
(which expresses the deviation from extremality). Considering first
the derived triplet, we obtained analytical expressions for two of
its members: $\Lambda\propto\Delta^{4}$ and $\left\langle T_{uu}^{-}\right\rangle _{\text{ren}}^{H-U}\propto\Delta^{6}$
(see Eqs. (\ref{eq:Lambda_delta-2}) and (\ref{eq:Tuudif_delta})
respectively). For the third member we got a numerical result: $\left\langle T_{uu}^{-}\right\rangle _{\text{ren}}^{H}\propto\Delta^{5}$
(see Eq. (\ref{eq:TuuH_delta})). Our analytical results (including
both the leading-order powers of $\Delta$ and the corresponding prefactors)
agree with the behavior seen in the numerically computed quantities,
as portrayed in Fig. \ref{Fig:derived_triplet}.

From these results we could easily obtain the leading-order behavior
of the elementary triplet, namely the flux quantities $\left\langle T_{yy}^{-}\right\rangle _{\text{ren}}^{\Xi}$
(see Subsec. \ref{subsec:elem_triplet}). We quote our final results:
\[
\left\langle T_{vv}^{-}\right\rangle _{\text{ren}}^{H}=\left\langle T_{uu}^{-}\right\rangle _{\text{ren}}^{H}\cong\left\langle T_{uu}^{-}\right\rangle _{\text{ren}}^{U}\cong\alpha\Delta^{5}\,\,\,,\,\,\,\left\langle T_{vv}^{-}\right\rangle _{\text{ren}}^{U}\cong-\hbar\frac{\Delta^{4}}{480\pi^{2}M^{4}}
\]
where $\alpha$ is a coefficient extracted from the numerics to be
$\alpha\cong-3.4375\times10^{-3}\,\hbar M^{-4}$ (as indicated from
the level of the horizontal dashed green line in e.g. Fig. \ref{Fig:derived_triplet}).

These results may be intuitively understood as follows. At a nearly
extremal RN BH, since the interior domain shrinks as the two horizons
``approach one another'' (as indicated by the similarity of their
$r_{+}$ and $r_{-}$ values, which only differ by $2M\Delta$), the
fluxes at the IH vicinity don't differ much from their corresponding
EH values. That is, since for an evaporating BH (in the Unruh state)
we have $\left\langle T_{vv}\right\rangle _{\text{ren}}^{U}<0$ and
$\left\langle T_{uu}\right\rangle _{\text{ren}}^{U}=0$ at the EH,
we expect to find at the IH a negative $\left\langle T_{vv}^{-}\right\rangle _{\text{ren}}^{U}$
(similar in magnitude to its corresponding EH value), as well as $\left\langle T_{uu}^{-}\right\rangle _{\text{ren}}^{U}$
vanishing more rapidly than $\left\langle T_{vv}^{-}\right\rangle _{\text{ren}}^{U}$,
as extremality is approached. In particular, this means the quantity
$\Lambda$ is expected to be dominated by $\left\langle T_{vv}^{-}\right\rangle _{\text{ren}}^{U}$,
and indeed we find the following approximate relation to hold near
extremality: $\Lambda\cong-4\pi r_{-}^{2}\left\langle T_{vv}^{-}\right\rangle _{\text{ren}}^{U}$
(see Subsec. \ref{subsec:elem_triplet}). 

Although our main interest in this paper concerns semiclassical physics
deep inside the BH, in passing, we also derived the leading-order
small-$\omega$ expression for $\tau_{\omega l}^{\text{up}}$, namely
the transmission coefficient \emph{outside} the BH (see Appendix).
This coefficient is a necessary ingredient in the analysis of the
near-IH flux differences $\left\langle T_{uu}^{-}\right\rangle _{\text{ren}}^{H-U}$
and $\Lambda$. 

As was already mentioned, the quantity $4\pi r^{2}\left(\left\langle T_{uu}\right\rangle _{\text{ren}}-\left\langle T_{vv}\right\rangle _{\text{ren}}\right)$
is independent of $r$ in both HH and Unruh states. In the latter,
at the IH it reduces to $\Lambda$ (given in Eq. (\ref{eq:Lambda_delta-2}))
whereas in the limit $r\to\infty$ it coincides with the Hawking-evaporation
outflux. Thus, on passing we have obtained the explicit expression
for the evaporation rate of a near-extremal RN BH: 
\begin{equation}
\lim_{r\to\infty}4\pi r^{2}\left\langle T_{uu}^{U}\right\rangle _{\text{ren}}\cong\hbar\,\frac{\Delta^{4}}{120\pi M^{2}}\,.\label{eq: Outflux}
\end{equation}
 While the scaling of this quantity as $\propto\Delta^{4}$ has already
been pointed out in e.g. Ref. \citep{Jacobson}, we are not aware
of previous derivations of the prefactor. The analytical computation
of this prefactor, carried out in Subsec. \ref{subsec:Lambda_analysis},
was made possible due to our analysis of the transmission coefficient
$\tau_{\omega l}^{\text{up}}$ at low frequencies (presented in the
Appendix). 

Returning to semiclassical fluxes inside the BH, our results indicate
that for a near-extremal RN BH in the Unruh state, $\left\langle T_{vv}\right\rangle _{\text{ren}}^{U}$
dominates over $\left\langle T_{uu}\right\rangle _{\text{ren}}^{U}$
in the IH vicinity. This could suggest that the semiclassically back-reacted
geometry in this domain may be well approximated by the ingoing charged
Vaidya solution \citep{BonnorVaidya:1970}. We hope to further explore
this issue in future research.

\appendix

\section{The transmission coefficient in low frequencies.}

In this appendix we analyze the leading order of the transmission
coefficient $\tau_{\omega l}$ at low frequencies (namely, corresponding
to modes with $\omega M\ll1$), in a RN BH. We shall provide the full
analysis for a subextremal BH (that is, with $Q/M<1$), which is of
direct relevance for this paper, and then quote the analogous result
for an extremal RN BH (with $Q/M=1$).

We shall consider the ``in'' mode normalized to have amplitude $1$
at the EH, denoted by $\hat{\psi}_{\omega l}$, which is a solution
to the radial equation (\ref{eq:rad_eq}) in the BH exterior with
the following asymptotic behavior:
\begin{equation}
\hat{\psi}_{\omega l}(r_{*})\cong\begin{cases}
e^{-i\omega r_{*}} & r_{*}\to-\infty\\
\mathcal{T}_{\omega l}e^{-i\omega r_{*}}+\mathcal{R}_{\omega l}e^{i\omega r_{*}} & r_{*}\to\infty
\end{cases}\,.\label{eq:in_BCs}
\end{equation}

$\mathcal{T}_{\omega l}$ and $\mathcal{R}_{\omega l}$ may then be
used to construct the usual reflection and transmission coefficients
$\tau_{\omega l}$ and $\rho_{\omega l}$ of the standard ``in''
and ``up'' Eddington-Finkelstein modes (see Ref. \citep{Group:2018}).
In the ``in'' modes, $\tau_{\omega l}^{\text{in}}$ and $\rho_{\omega l}^{\text{in}}$
are trivially related to $\mathcal{T}_{\omega l}$ and $\mathcal{R}_{\omega l}$
via:
\begin{align}
\tau_{\omega l}^{\text{in}} & =\frac{1}{\mathcal{T}_{\omega l}}\,\;,\,\;\rho_{\omega l}^{\text{in}}=\frac{\mathcal{R}_{\omega l}}{\mathcal{T}_{\omega l}}\,.\label{eq:tau_rho_relations_in}
\end{align}
The corresponding ``up'' mode coefficients, $\tau_{\omega l}^{\text{up}}$
and $\rho_{\omega l}^{\text{up}}$, may be related to their ``in''
counterparts through the conserved Wronskian, yielding:
\begin{equation}
\tau_{\omega l}^{\text{up}}=\tau_{\omega l}^{\text{in}}=\frac{1}{\mathcal{T}_{\omega l}}\,\;,\,\;\rho_{\omega l}^{\text{up}}=-\rho_{\omega l}^{\text{in}*}\frac{\tau_{\omega l}^{\text{in}}}{\tau_{\omega l}^{*\text{in}}}=-\frac{\mathcal{R}_{\omega l}^{*}}{\mathcal{T}_{\omega l}}\,.\label{eq:tau_rho_relations_up}
\end{equation}
We denote $\tau_{\omega l}\equiv\tau_{\omega l}^{\text{in}}=\tau_{\omega l}^{\text{up}}$,
and we take here the $r_{*}$ convention used in Ref. \citep{Group:2018}:
\footnote{Note that the result for $\tau_{\omega l}$ is independent of the
choice of integration constant for $r_{*}$. With a different choice,
say $\tilde{r}_{*}\equiv r_{*}+\delta r_{*}$ ($\delta r_{*}$ being
a constant), the desired asymptotic behavior at the EH will now naturally
be $e^{-i\omega\tilde{r}_{*}}$, which amounts to multiplying Eq.
(\ref{eq:in_BCs}) by the constant phase $e^{-i\omega\delta r_{*}}$.
This yields 
\[
\hat{\psi}_{\omega l}(r_{*})\cong\begin{cases}
e^{-i\omega\tilde{r}_{*}} & r_{*}\to-\infty\\
\mathcal{T}_{\omega l}e^{-i\omega\tilde{r}_{*}}+\left(\mathcal{R}_{\omega l}e^{-2i\omega\delta r_{*}}\right)e^{i\omega\tilde{r}_{*}} & r_{*}\to\infty
\end{cases}\,\,.
\]
That is, $\mathcal{T}_{\omega l}$ hasn't changed and hence, from
Eq. (\ref{eq:tau_rho_relations_up}), $\tau_{\omega l}$ is left unaffected.
On the other hand, $\mathcal{R}_{\omega l}$ has gained a phase of
$e^{-2i\omega\delta r_{*}}$, which translates to the same effect
on $\rho_{\omega l}$. Nevertheless, it is not difficult to show that
the leading order of $\rho_{\omega l}$ at small $\omega$ (given
below in Eq. (\ref{eq:rho_smallw})) remains unaffected.}
\begin{align}
r_{*} & =r+\frac{1}{2\kappa_{+}}\log\left(\frac{r-r_{+}}{r_{+}-r_{-}}\right)-\frac{1}{2\kappa_{-}}\log\left(\frac{r-r_{-}}{r_{+}-r_{-}}\right)\,.\label{eq:rstar}
\end{align}

The variation of the effective potential $V_{l}$ given in Eq. (\ref{eq:eff_pot})
between the EH ($r_{*}\to-\infty$) and infinity ($r_{*}\to\infty$)
suggests a natural division of the BH exterior into three overlapping
regions, in which suitable approximations can be made: region I at
the EH vicinity, where the effect of the potential is negligible and
we may approximate the radial function by a free solution $\psi\cong e^{-i\omega r_{*}}$
(a more detailed characterization of this region will follow); region
II where $\omega$ is negligible in the radial equation, that is,
the domain characterized by $\omega^{2}\ll V_{l}$; and region III,
the asymptotically flat region, where $r/M\gg1$. Note that due to
our assumption of small frequencies region II is very vast and, as
we shall see, indeed overlaps with both its neighboring regions. This,
in turn, allows the matching procedure which follows, relating the
asymptotic regions $r_{*}\to-\infty$ and $r_{*}\to\infty$. We shall
start at the EH vicinity and work our way outwards to infinity, where
the reflection and transmission coefficients are to be extracted.

\subsubsection{Region I}

We start our analysis at the asymptotic domain where the effective
potential is negligible, satisfying $V_{l}\ll\omega^{2}$. This yields
a free solution to the radial equation (\ref{eq:rad_eq}), which according
to Eq. (\ref{eq:in_BCs}) is
\begin{equation}
\psi_{\omega l}^{\text{free}}=e^{-i\omega r_{*}}\,.\label{eq:free_sol}
\end{equation}
However, the domain characterized by $V_{l}\ll\omega^{2}$ has no
overlap with region II, where, as mentioned above, $V_{l}\gg\omega^{2}$.
We thus wish to ``enhance'' the free solution $\psi_{\omega l}^{\text{free}}$,
in order to slightly extend its domain of validity. To build the \emph{enhanced
free solution}, we consider the leading order near-EH form of the
potential, $V_{l}\cong v_{l}M^{-2}\exp\left(2\kappa_{+}r_{*}\right)$,
where $v_{l}$ is a certain dimensionless constant \footnote{Note that $V_{l}\propto f(r)\propto r-r_{+}$ in the EH vicinity,
and then evaluating Eq. (\ref{eq:rstar}) at $r\cong r_{+}$ yields
the relation to $r_{*}$, namely $r-r_{+}\propto\exp\left(2\kappa_{+}r_{*}\right)$.}. Correspondingly, we use the Ansatz 
\[
\psi_{\omega l}\cong\psi_{\omega l}^{\text{free}}\left[1+cM^{2}V_{l}\right]\cong e^{-i\omega r_{*}}\left[1+cv_{l}\exp\left(2\kappa_{+}r_{*}\right)\right]\,,
\]
where $c$ is a dimensionless constant that will be determined by
the radial equation as follows: Applying the differential operator
$\text{d}^{2}/\text{d}r_{*}^{2}+(\omega^{2}-V_{l})$ to this Ansatz
for $\psi_{\omega l}$ yields 
\[
\left[\text{d}^{2}/\text{d}r_{*}^{2}+(\omega^{2}-V_{l})\right]\psi_{\omega l}=e^{-i\omega r_{*}}V_{l}\left[4M^{2}\kappa_{+}\left(\kappa_{+}-i\omega\right)c-1\right]+\mathcal{O}(V_{l}^{2})\,.
\]
Equating the right hand side to zero (ignoring the $\mathcal{O}(V_{l}^{2})$
term) yields $c=\left[4M^{2}\kappa_{+}\left(\kappa_{+}-i\omega\right)\right]^{-1}$.
Thus, we find the near-EH solution (to leading order in $V_{l}$)
to be 
\begin{equation}
\psi_{\omega l}=e^{-i\omega r_{*}}\left[1+\frac{1}{4\kappa_{+}\left(\kappa_{+}-i\omega\right)}V_{l}\right]\,.\label{eq:psi_enhanced}
\end{equation}
For later convenience, we also write its derivative with respect to
$r_{*}$ (hereafter denoted by a prime):
\begin{equation}
\psi'_{\omega l}=-i\omega e^{-i\omega r_{*}}\left[1-\frac{2\kappa_{+}-i\omega}{4i\omega\kappa_{+}\left(\kappa_{+}-i\omega\right)}V_{l}\right]\,.\label{eq:dpsi_enhanced}
\end{equation}

The domain of validity of this approximation (which ignores terms
of higher orders in $V_{l}$) is basically characterized by $M^{2}V_{l}\ll1$.
However, for our goal of subsequently matching this solution to region
II, it will be convenient to further restrict region I such that both
$\psi_{\omega l}$ and $\psi'_{\omega l}$ are still not significantly
affected by the potential $V_{l}$. (We are concerned about the forms
of $\psi_{\omega l}$ as well as $\psi'_{\omega l}$, because the
matching of regions I and II will involve the values of both $\psi_{\omega l}$
and $\psi'_{\omega l}$ in the overlap domain). Recalling that $\kappa_{+}\sim1/M\gg\omega$,
one readily sees that the more stringent restriction emerges from
the expression for $\psi'_{\omega l}$: The term in squared brackets
in Eq. (\ref{eq:dpsi_enhanced}) reads $\approx1-V_{l}/(2i\omega\kappa_{+})$
for small $\omega$, hence the demand that $\psi'_{\omega l}$ remains
well approximated by its free counterpart $\psi'{}_{\omega l}^{\text{free}}=-i\omega e^{-i\omega r_{*}}$
yields the requirement

\begin{equation}
V_{l}\left(r\right)\ll\omega/M\,,\qquad r\approx r_{+}\qquad\qquad\qquad\text{(region I)}\label{eq:con_I}
\end{equation}
 \footnote{The restriction $r\approx r_{+}$ was added in this equation to indicate
that, obviously, it is only the small-$V_{l}$ domain \emph{at the
EH vicinity} (and not the one at $r\gg M$) that defines region I.
The same remark also applies to Eq. (\ref{eq:overlap_I-II}) below.}. The last inequality guarantees that both $\psi'_{\omega l}$ and
$\psi_{\omega l}$ do not differ much from their free values $\psi'{}_{\omega l}^{\text{free}}$
and $\psi_{\omega l}^{\text{free}}$. We thus take this inequality
to characterize region I, and we denote the approximate solution therein
by $\hat{\psi}_{\omega l}^{I}$. From the very construction of region
I, we may simply take $\hat{\psi}_{\omega l}^{I}$ to be the free
solution given in Eq. (\ref{eq:free_sol}) \footnote{The fact that $\hat{\psi}_{\omega l}^{I}$ and its derivative attain
values similar to their free-solution counterparts throughout the
domain (\ref{eq:con_I}) may seem surprising at first sight, because
the potential $V_{l}\left(r\right)$ is not negligible compared to
$\omega^{2}$ everywhere throughout that domain (in fact, we even
have $V_{l}\gg\omega^{2}$ in some portion of the latter). The reason
for this similarity is simple: The width of the sub-domain where $V_{l}\left(r\right)$
fails to be $\ll\omega^{2}$ is merely of order $M$; and even in
this sub-domain $V_{l}$ is still $\ll\omega/M$. It therefore follows
that $\hat{\psi}_{\omega l}^{I}$ and its derivative do not accumulate
a significant deviation from their corresponding free values along
that limited sub-domain. }.

Having $V_{l}\propto r-r_{+}$ in that domain, we may rewrite the
condition in Eq. (\ref{eq:con_I}) as
\begin{equation}
\frac{r-r_{+}}{M}\ll\omega M\,.\qquad\qquad\qquad\text{(region I)}\label{eq:con_I_dr}
\end{equation}
Note that since $\omega M\ll1$, the last inequality also ensures
that region I is indeed at the EH vicinity, where the assumed near-EH
form of the potential is valid. 

\subsubsection{Region II}

This region is characterized by 
\begin{equation}
V_{l}\left(r\right)\gg\omega^{2}\,,\qquad\qquad\qquad\text{(region II)}\label{eq:con_II}
\end{equation}
and we may thus neglect $\omega^{2}$ in the radial equation (\ref{eq:rad_eq})
as a leading order approximation. This yields the so-called \emph{static
solution},

\begin{equation}
\hat{\psi}_{\omega l}^{II}=\frac{r}{M}\left[C_{1}P_{l}\left(\frac{r-M}{r_{+}-M}\right)+C_{2}Q_{l}\left(\frac{r-M}{r_{+}-M}\right)\right]\,,\label{eq:sol_II}
\end{equation}
where $P_{l}$ and $Q_{l}$ are respectively the Legendre polynomial
and Legendre function of the second kind \footnote{$Q_{l}\left(x\right)$ is defined here as the real branch in the domain
$x>1$ (corresponding here to $r>r_{+}$, namely, the BH exterior).
This function is classified in \emph{Wolfram Mathematica} as the ``Legendre
function of type 3''.}, and $C_{1},C_{2}$ are coefficients to be determined. We shall treat
$\hat{\psi}_{\omega l}^{II}$ as the approximate solution throughout
region II.

Owing to the basic assumption of low frequencies $\omega M\ll1$,
region II (characterized in Eq. (\ref{eq:con_II})) and region I (characterized
in Eq. (\ref{eq:con_I})) overlap in a domain satisfying
\begin{equation}
\left(\omega M\right)^{2}\ll V_{l}M^{2}\ll\omega M\,,\qquad r\approx r_{+}\qquad\qquad\qquad\text{(regions I-II overlap)}\label{eq:overlap_I-II}
\end{equation}
or, from the near-EH form of $V_{l}$, 
\begin{equation}
\left(\omega M\right)^{2}\ll\frac{r-r_{+}}{M}\ll\omega M\,.\qquad\qquad\qquad\text{(regions I-II overlap)}\label{eq:overlap_I-II_dr}
\end{equation}

In order to match the solutions $\hat{\psi}_{\omega l}^{I}$ and $\hat{\psi}_{\omega l}^{II}$
in the overlap domain characterized above, we apply the right side
of the inequality (\ref{eq:overlap_I-II_dr}) in $\hat{\psi}_{\omega l}^{II}$
and the left side of this inequality in $\hat{\psi}_{\omega l}^{I}$.
In fact, it turns out to be sufficient (and equivalent) to take $r-r_{+}\ll M$
in $\hat{\psi}_{\omega l}^{II}$ and $|\omega r_{*}|\ll1$ in $\hat{\psi}_{\omega l}^{I}$
\footnote{Note that in the EH-vicinity, setting $r\cong r_{+}$ in Eq. (\ref{eq:rstar})
yields $r_{*}\cong r_{+}+\frac{1}{2\kappa_{+}}\log\left(\frac{r-r_{+}}{r_{+}-r_{-}}\right)\sim M\log\left(\frac{r-r_{+}}{M}\right)$.
Then, choosing a typical point in the overlap domain (\ref{eq:overlap_I-II_dr}),
e.g. $\frac{r-r_{+}}{M}\sim\left(\omega M\right)^{\gamma}$ for some
fixed positive $\gamma$ (noting that this overlap domain actually
corresponds to $1<\gamma<2$), we have $|\omega r_{*}|\sim\gamma\omega M|\log\left(\omega M\right)|$,
which is $\ll1$ due to the basic assumption of low frequencies. That
is, the condition $|\omega r_{*}|\ll1$ is guaranteed to hold throughout
the overlap domain (\ref{eq:overlap_I-II_dr}).}.

Taking the solution $\hat{\psi}_{\omega l}^{I}$ given in Eq. (\ref{eq:free_sol})
in the asymptotic domain of region I where $|\omega r_{*}|\ll1$ yields:

\begin{align}
\hat{\psi}_{\omega l}^{I}\left(|\omega r_{*}|\ll1\right) & \cong1-i\omega r_{*}\cong1-i\omega\left[r_{+}+\frac{r_{+}^{2}}{r_{+}-r_{-}}\log\left(\frac{r-r_{+}}{r_{+}-r_{-}}\right)\right]\label{eq:sol_I_asyR}\\
\frac{\text{d}}{\text{d}r}\hat{\psi}_{\omega l}^{I}\left(|\omega r_{*}|\ll1\right) & \cong-i\omega\frac{r_{+}^{2}}{\left(r-r_{+}\right)\left(r_{+}-r_{-}\right)}\,.\nonumber 
\end{align}

Carrying the solution $\hat{\psi}_{\omega l}^{II}$ as given in Eq.
(\ref{eq:sol_II}) to the asymptotic domain of region II where $r-r_{+}\ll M$,
and using the leading-order asymptotic behavior of our basis functions
$P_{l}\left(x\to1^{+}\right)=1$ and $Q_{l}\left(x\to1^{+}\right)\cong\frac{1}{2}\ln\left(\frac{2}{x-1}\right)$
\footnote{In fact, $Q_{l}\left(x\to1^{+}\right)\approx\frac{1}{2}\ln\left(\frac{2}{x-1}\right)-h\left(l\right)$,
but we may neglect the constant $h\left(l\right)$ compared to the
logarithmically diverging term. (We should also note that this ``parasitic''
constant does not interfere with the extraction of $C_{1}$ from the
first equation in (\ref{eq:sol_II_asyL}), because $C_{2}$ turns
out to be $\propto\omega$, hence $C_{1}$ is determined right away
from the $\omega$-independent part of Eq. (\ref{eq:sol_I_asyR}).)}, we get
\begin{align}
\hat{\psi}_{\omega l}^{II}\left(r-r_{+}\ll M\right) & \cong\frac{r_{+}}{M}\left[C_{1}+\frac{1}{2}C_{2}\log\left(\frac{r_{+}-r_{-}}{r-r_{+}}\right)\right]\label{eq:sol_II_asyL}\\
\frac{\text{d}}{\text{d}r}\hat{\psi}_{\omega l}^{II}\left(r-r_{+}\ll M\right) & \cong-\frac{1}{2M}\frac{r_{+}}{r-r_{+}}C_{2}\nonumber 
\end{align}
regardless of $l$. Then, matching to Eq. (\ref{eq:sol_I_asyR}) requires
setting the coefficients $C_{1},C_{2}$ to their leading order in
$\omega$ (which suffices for the present analysis) as follows:
\[
C_{1}=\frac{M}{r_{+}}{\color{blue}}\,\,,\,\,C_{2}=2i\omega M\frac{r_{+}}{r_{+}-r_{-}}\,.
\]
Feeding this in Eq. (\ref{eq:sol_II}), the approximate solution in
region II is found to be:
\begin{equation}
\hat{\psi}_{\omega l}^{II}=\frac{r}{r_{+}}P_{l}\left(\frac{r-M}{r_{+}-M}\right)+\frac{2i\omega r_{+}r}{r_{+}-r_{-}}Q_{l}\left(\frac{r-M}{r_{+}-M}\right)\,.\label{eq:sol_II_fin}
\end{equation}

Finally, we explore the domain of validity of the region-II approximation
in the range $r\gg M$. The basic criterion that needs to be satisfied
in this region is given in Eq. (\ref{eq:con_II}), namely $V_{l}\left(r\right)\gg\omega^{2}$.
At $r\gg M$, the effective potential $V_{l}$ given in Eq. (\ref{eq:eff_pot})
decays like $\propto1/r^{2}$ for $l>0$ and like $\propto M/r^{3}$
for $l=0$. This implies that the corresponding domain of validity
is $r/M\ll\left(\omega M\right)^{-1}$ for $l>0$ and $r/M\ll\left(\omega M\right)^{-2/3}$
for $l=0$. In the analysis that follows it will be convenient to
treat the $l=0$ and $l>0$ cases on a common footing. We therefore
choose the domain in which we apply the region-II approximation, in
the range $r\gg M$, to be the stringent of these two domains (that
is, the one emerging from the $l=0$ case):
\begin{equation}
r/M\ll\left(\omega M\right)^{-2/3}\,.\qquad\qquad\qquad\text{(region II, large-\ensuremath{r} side)}\label{eq: Large-r_II}
\end{equation}

\subsubsection{Region III}

In the asymptotically flat region characterized by
\begin{equation}
r/M\gg1\qquad\qquad\qquad\text{(region III)}\label{eq:con_III}
\end{equation}
(which implies $f\left(r\right)\cong1$, $\text{d}f/\text{d}r\cong0$),
the approximate solution is well known and is given in terms of spherical
Bessel functions:

\begin{align}
\hat{\psi}_{\omega l}^{III} & =\omega r_{*}\left[D_{1}j_{l}\left(\omega r_{*}\right)+D_{2}y_{l}\left(\omega r_{*}\right)\right]\,,\label{eq:sol_III}
\end{align}
where $j_{l}$ and $y_{l}$ are respectively the spherical Bessel
functions of the first and second kind, and $D_{1},D_{2}$ are coefficients
to be determined from the matching procedure. \footnote{In principle one could also write down another approximate solution
$\tilde{\psi}_{\omega l}^{III}$ in this $r/M\gg1$ region, which
takes the same form as $\hat{\psi}_{\omega l}^{III}$ but with $r_{*}$
replaced by $r$. A direct inspection indicates, however, that the
error involved in $\tilde{\psi}_{\omega l}^{III}$ is much larger
than that involved in $\hat{\psi}_{\omega l}^{III}$. To see this,
one can substitute these approximate solutions in the radial equation
(\ref{eq:rad_eq}). The (relative) error is then found to scale as
$\propto M/r$ for $\tilde{\psi}_{\omega l}^{III}$, and only $\propto l\left(l+1\right)(M/r)^{3}\ln(r/M)$
(or $\propto(M/r)^{3}$ in the $l=0$ case) for $\hat{\psi}_{\omega l}^{III}$.
In fact, this larger error in $\tilde{\psi}_{\omega l}^{III}$ is
manifested, at the large-$r$ limit, in the phase that erroneously
progresses in this solution like $\omega r$ instead of $\omega r_{*}$.
(Also recall that the difference $r_{*}-r$ actually diverges logarithmically
at large $r$. Therefore $\tilde{\psi}_{\omega l}^{III}$ fails to
be a valid approximate solution in a global sense, even at arbitrarily
large $r$.)}

We wish to match $\hat{\psi}_{\omega l}^{III}$ with the solution
$\hat{\psi}_{\omega l}^{II}$ of region II. The overlap domain of
regions II and III is obtained by combining the conditions (\ref{eq: Large-r_II})
and (\ref{eq:con_III}), namely:
\begin{equation}
1\ll\frac{r}{M}\ll\left(\omega M\right)^{-2/3}\,.\qquad\qquad\qquad\text{(regions II-III overlap)}\label{eq:overlap_II-III}
\end{equation}
This overlap domain indeed exists, owing to our basic assumption $\omega M\ll1$.
\footnote{Note that we may replace $r$ in Eq. (\ref{eq:overlap_II-III}) by
$r_{*}$, leaving the inequality unaffected. This follows from the
simple fact that $r_{*}/r\cong1$ throughout the domain $r\gg M$.} Furthermore, a direct consequence of Eq. (\ref{eq:overlap_II-III})
is $r/M\ll\left(\omega M\right)^{-1}$ and therefore:
\begin{equation}
\omega r\ll1\,.\label{eq:con_II_R}
\end{equation}
We find it convenient to describe the matching in the overlap domain
to be between $\hat{\psi}_{\omega l}^{III}$ in the asymptotic domain
of region III where $\omega r_{*}\ll1$, and $\hat{\psi}_{\omega l}^{II}$
in the asymptotic domain of region II where $r/M\gg1$. The large-$r$
limit of $\hat{\psi}_{\omega l}^{II}$ (given in Eq. (\ref{eq:sol_II_fin}))
is obtained from the asymptotic behavior of $P_{l}\left(x\right)$
and $Q_{l}\left(x\right)$ at a large argument, namely $P_{l}\left(x\to\infty\right)\cong\left(2x\right)^{l}\frac{\Gamma\left(l+\frac{1}{2}\right)}{l!\sqrt{\pi}}$
and $Q_{l}\left(x\to\infty\right)\cong\left(2x\right)^{-l-1}\frac{l!\sqrt{\pi}}{\Gamma\left(l+\frac{3}{2}\right)}$.
Inserting that into Eq. (\ref{eq:sol_II_fin}) yields:
\begin{equation}
\hat{\psi}_{\omega l}^{II}\left(r/M\gg1\right)\cong r^{l+1}\frac{1}{r_{+}}\left(\frac{2}{r_{+}-M}\right)^{l}\frac{\Gamma\left(l+\frac{1}{2}\right)}{\sqrt{\pi}l!}+r^{-l}\frac{2i\omega r_{+}}{r_{+}-r_{-}}\frac{\sqrt{\pi}l!}{\Gamma\left(l+\frac{3}{2}\right)}\left(\frac{2}{r_{+}-M}\right)^{-l-1}\,.\label{eq:sol_II_asyR}
\end{equation}
Plugging the asymptotic behavior of the spherical Bessel functions
of the first and second kinds at a small argument in Eq. (\ref{eq:sol_III}),
we obtain:
\begin{equation}
\hat{\psi}_{\omega l}^{III}\left(\omega r_{*}\ll1\right)\cong\omega r_{*}\left[D_{1}\left(\omega r_{*}\right)^{l}\frac{\sqrt{\pi}}{\Gamma\left(\frac{3}{2}+l\right)}2^{-l-1}-D_{2}\left(\omega r_{*}\right)^{-l-1}\frac{1}{\sqrt{\pi}}2^{l}\Gamma\left(l+\frac{1}{2}\right)\right]\,,\label{eq:sol_III_asyL}
\end{equation}
Note that the dependence on $r$ in the last two equations is only
through simple powers of $r$ or $r_{*}$. We can then re-express
these two equations in the more compact form
\[
\hat{\psi}_{\omega l}^{III}\left(\omega r_{*}\ll1\right)\cong\tilde{D}_{1}r_{*}^{l+1}+\tilde{D}_{2}r_{*}^{-l}
\]
\[
\hat{\psi}_{\omega l}^{II}\left(r/M\gg1\right)\cong\tilde{C}_{1}r^{l+1}+\tilde{C}_{2}r^{-l}
\]
(where the new coefficients $\tilde{C}_{i},\tilde{D}_{i}$ are trivially
related to $C_{i},D_{i}$ by comparing the above to Eqs. (\ref{eq:sol_II_asyR},\ref{eq:sol_III_asyL})).
Obviously, the large-$r$ assumption allows replacing $r_{*}^{l+1}$
with $r^{l+1}$ and $r_{*}^{-l}$ with $r^{-l}$, as the relative
error decays like $\propto\frac{M}{r}\log\left(\frac{r}{M}\right)\ll1$.
The matching then simply yields $\tilde{D}_{1}=\tilde{C}_{1}$ and
$\tilde{D}_{2}=\tilde{C}_{2}$. Applying this straightforward matching
scheme to Eqs. (\ref{eq:sol_II_asyR},\ref{eq:sol_III_asyL}) determines
the desired coefficients $D_{1},D_{2}$ (to their leading order in
$\omega$):
\begin{align}
D_{1}=\lambda_{l}\left(\omega M\right)^{-l-1}\,\,,\,\,D_{2}=-i\lambda_{l}^{-1}\left(\omega M\right)^{l+1}\label{eq:D1,2}
\end{align}
where
\begin{equation}
\lambda_{l}=\frac{M}{r_{+}}\left(\frac{8M}{r_{+}-r_{-}}\right)^{l}\frac{2\Gamma\left(l+\frac{1}{2}\right)\Gamma\left(l+\frac{3}{2}\right)}{\pi l!}\,.\label{eq:lambda_l_const}
\end{equation}

Feeding Eq. (\ref{eq:D1,2}) into Eq. (\ref{eq:sol_III}), we obtain
the approximate solution throughout region III: 
\begin{align}
\hat{\psi}_{\omega l}^{III} & =\omega r_{*}\left[\lambda_{l}\left(\omega M\right)^{-l-1}j_{l}\left(\omega r_{*}\right)-i\lambda_{l}^{-1}\left(\omega M\right)^{l+1}y_{l}\left(\omega r_{*}\right)\right]\,.\label{eq:sol_III-fin}
\end{align}

\subsubsection{Asymptotic behavior at $r\to\infty$}

Finally, in order to extract $\mathcal{T}_{\omega l}$ and $\mathcal{R}_{\omega l}$,
we need to match $\hat{\psi}_{\omega l}^{III}$ to the boundary condition
(\ref{eq:in_BCs}) at $r_{*}\to\infty$. That is, we are interested
in the asymptotic behavior of $\hat{\psi}_{\omega l}^{III}$ where
$\omega r_{*}\gg1$. Using $j_{l}\left(x\to\infty\right)\cong-\frac{1}{x}\sin\left(\frac{l\pi}{2}-x\right)$
and $y_{l}\left(x\to\infty\right)\cong-\frac{1}{x}\cos\left(\frac{l\pi}{2}-x\right)$
in Eq. (\ref{eq:sol_III}), we obtain:
\begin{equation}
\hat{\psi}_{\omega l}^{III}\left(\omega r_{*}\gg1\right)\cong-D_{1}\sin\left(\frac{l\pi}{2}-\omega r_{*}\right)-D_{2}\cos\left(\frac{l\pi}{2}-\omega r_{*}\right)\,.\label{eq:sol_III_asyR}
\end{equation}

At the $\omega M\ll1$ limit, the coefficient $D_{2}\propto\left(\omega M\right)^{l+1}$
is negligible compared to $D_{1}\propto\left(\omega M\right)^{-l-1}$
(see Eq. (\ref{eq:D1,2})), and we are left with:
\begin{align}
\hat{\psi}_{\omega l}^{III}\left(\omega r_{*}\gg1\right) & \cong-\lambda_{l}\left(\omega M\right)^{-l-1}\sin\left(\frac{l\pi}{2}-\omega r_{*}\right)\nonumber \\
 & =\frac{\lambda_{l}}{2}\left(\omega M\right)^{-l-1}i^{l+1}\left[e^{-i\omega r_{*}}+\left(-1\right)^{l+1}e^{i\omega r_{*}}\right]\,.\label{eq:asym_psi_subext}
\end{align}

With the above asymptotic form, we can easily read the coefficients
$\mathcal{T}_{\omega l}$ and $\mathcal{R}_{\omega l}$ as appearing
in Eq. (\ref{eq:in_BCs}):
\[
\mathcal{T}_{\omega l}=\frac{\lambda_{l}}{2}\left(\omega M\right)^{-l-1}i^{l+1}\,\,\,\,,\,\,\,\,\mathcal{R}_{\omega l}=\frac{\lambda_{l}}{2}\left(\omega M\right)^{-l-1}\left(-1\right)^{l+1}i^{l+1}\,.
\]
Then, via the relations in Eqs. (\ref{eq:tau_rho_relations_in},\ref{eq:tau_rho_relations_up}),
one can readily extract the reflection and transmission coefficients
to leading order in low frequencies:
\begin{equation}
\rho_{\omega l}^{\text{in}}\cong\left(-1\right)^{l+1}\,\,,\,\,\rho_{\omega l}^{\text{up}}\cong-1\label{eq:rho_smallw}
\end{equation}
and

\begin{equation}
\tau_{\omega l}\cong\frac{\pi r_{+}}{M}\left(\frac{r_{+}-r_{-}}{8M}\right)^{l}\frac{l!}{\Gamma\left(l+\frac{1}{2}\right)\Gamma\left(l+\frac{3}{2}\right)}\left(-i\right)^{l+1}\left(\omega M\right)^{l+1}\,,\label{eq:tau_smallw}
\end{equation}
or, using $\Gamma\left(\frac{1}{2}+l\right)=\frac{\left(2l\right)!}{4^{l}l!}\sqrt{\pi}$:
\begin{equation}
\tau_{\omega l}\cong\frac{r_{+}}{M}\left(\frac{r_{+}-r_{-}}{M}\right)^{l}\frac{2^{l+2}\left(l!\right)^{2}\left(l+1\right)!}{\left(2l\right)!\left(2l+2\right)!}\left(-i\right)^{l+1}\left(\omega M\right)^{l+1}\,.\label{eq:tau_final}
\end{equation}

One immediate consequence is that the leading order of $\tau_{\omega l}$
in small frequencies is real when $l$ is odd and imaginary when $l$
is even. In particular, for the sake of this paper, note that for
$l=0$ we have to leading order
\begin{equation}
\tau_{\omega,l=0}\cong-2i\omega r_{+}\,.\label{eq:tau_smallw-l0}
\end{equation}

The results presented here were verified numerically -- both for
$l=0$ as given in Eq. (\ref{eq:tau_smallw-l0}) and for several other
$l$ values as given more generally in Eq. (\ref{eq:tau_final}) --
in a variety of subextremal $Q/M$ values.

In the Schwarzschild limit ($r_{-}\to0$, $r_{+}\to2M$), Eq. (\ref{eq:tau_final})
adequately reduces to the corresponding result given in Eq. (5.5)
of Ref. \citep{CasalsOttewill:2015}. 

Note that the results presented in Eqs. (\ref{eq:rho_smallw},\ref{eq:tau_final})
were derived in the subextremal RN case only, and they are not valid
for an extremal BH. We shall briefly refer to the extremal case in
the subsection that follows.

\subsection*{The transmission coefficient in low frequencies in an extremal RN
BH}

An analysis analogous to the one presented in detail above can be
done in the extremal case. Since the two horizons now coincide at
$r=M$, this changes the behavior of $f\left(r\right)$, and hence
also $r_{*}$, $V_{l}\left(r\right)$, and the corresponding solutions
in the various domains. Nevertheless, despite these differences, the
basic strategy presented above is applicable in the extremal case
as well: We can again define the three domains with three corresponding
approximate solutions (the enhanced free solution in the EH vicinity,
the static solution where $V_{l}\gg\omega^{2}$, and the large-$r$
solution), with appropriate overlapping domains in which any two of
the neighboring approximate solutions may be matched. Then, matching
through and taking the $r_{*}\to\infty$ limit, we finally obtain
the asymptotic behavior (analogous to Eq. (\ref{eq:asym_psi_subext})
in the subextremal case): 
\begin{equation}
\hat{\psi}_{\omega l}^{III}\left(\omega r_{*}\gg1\right)\cong-\frac{i}{M}\left(\frac{2}{M}\right)^{2l}\Gamma\left(l+\frac{1}{2}\right)\Gamma\left(l+\frac{3}{2}\right)\frac{1}{\pi}\omega^{-2l-1}\left[\left(-1\right)^{l+1}e^{-i\omega r_{*}}+e^{i\omega r_{*}}\right]\,.\label{eq:asym_psi_ext}
\end{equation}
We may now use Eqs. (\ref{eq:tau_rho_relations_in},\ref{eq:tau_rho_relations_up})
to extract the transmission and reflection coefficients to leading
order in small frequencies for an extremal RN BH:
\begin{equation}
\rho_{\omega l}^{\text{in}}\cong\rho_{\omega l}^{\text{up}}\cong\left(-1\right)^{l+1}\label{eq:rho_smallw_ext}
\end{equation}
\begin{equation}
\tau_{\omega l}\cong i\left(-1\right)^{l+1}\frac{\pi}{2^{2l}}\frac{1}{\Gamma\left(l+\frac{1}{2}\right)\Gamma\left(l+\frac{3}{2}\right)}\left(\omega M\right)^{2l+1}\,.\label{eq:tau_smallw_ext}
\end{equation}

Note that the leading order of $\tau_{\omega l}$ in low frequencies
is $\propto\left(\omega M\right)^{2l+1}$ in the extremal case (unlike
$\left(\omega M\right)^{l+1}$ in the subextremal case, see Eq. (\ref{eq:tau_final})),
and that it is always imaginary.

\begin{acknowledgments}
We would like to thank Adam Levi for interesting discussions. This
work was supported by the Israel Science Foundation under Grant No.
600/18.
\end{acknowledgments}

\end{document}